\documentclass[journal, comsoc]{IEEEtran}

\usepackage{mathtools}

\DeclarePairedDelimiter\floor{\lfloor}{\rfloor}

\usepackage{dirtytalk}
\usepackage{times}
\usepackage{physics}
\usepackage{graphicx}
\usepackage[font=footnotesize]{caption}
\usepackage{amsmath}
\usepackage{amsfonts}
\usepackage{amssymb}
\usepackage{amsthm}
\usepackage{tabularx}
\usepackage{mathrsfs} 
\usepackage{times}
\usepackage{cite}

\usepackage[export]{adjustbox}
\usepackage{subfig}

\usepackage{soul}

\usepackage{stfloats}
\usepackage{float}
\usepackage{url}
\usepackage{hyperref}

\newtheorem{proposition}{Proposition}

\usepackage[nolist,nohyperlinks]{acronym}
\usepackage[numbers,sort&compress]{natbib}

\usepackage{bbm}

\usepackage{array}
\usepackage{booktabs}
\usepackage{colortbl}
\usepackage{mdwmath}
\usepackage{mdwtab}
\usepackage{eqparbox}
\usepackage{cellspace}
\usepackage{makecell}

\DeclareGraphicsExtensions{.png,.eps,.ps,.pdf}

\usepackage{tikz}
\usetikzlibrary{arrows,    
                shapes, 
                positioning, 
                fit,    
                backgrounds, 
                mindmap,
                petri,
                intersections, 
                external %
                }
\definecolor{emerald}{RGB}{69,155,61}
\definecolor{gold}{RGB}{244,216,51}
\definecolor{pink}{RGB}{235,44,206}
\usepackage{verbatim}

\tikzstyle{int}=[draw, fill=cyan!20, minimum size=2em]
\tikzstyle{int_blue}=[draw, fill=blue!20, minimum size=2em]
\tikzstyle{int_green}=[draw, fill=green!20, minimum size=2em]
\tikzstyle{int_red}=[draw, fill=red!20, minimum size=2em]
\tikzstyle{init} = [pin edge={to-,thin,black}]

\usepackage{svg}
\usepackage{filemod}
\usepackage{pgfplots}

\begin{document}

\title{Aging-Resistant Wideband Precoding in 5G and Beyond Using 3D Convolutional Neural Networks}



\author{Alejandro~Villena-Rodriguez,
        Francisco J. Mart\'in-Vega,
        Gerardo G\'omez,
        \\Mari Carmen Aguayo-Torres,
        and Georges Kaddoum,~\IEEEmembership{Senior Member, IEEE}
\thanks{Manuscript received April xx, 2024; revised XXX. %
This work has been sopported by Grant PID2022-137522OB-I00 funded by MCIN/AEI/10.13039/501100011033 and by FEDER A way of making Europe, Junta de Andalucia and University of Malaga (UMA) through grants PID2022-137522OB-I00 and RYC2021-034620-I. 
}
\thanks{Villena-Rodriguez, Mart\'in-Vega, G\'omez and Aguayo-Torres are with the Communications and Signal Processing Lab, Telecommunication Research Institute (TELMA), Universidad de M\'alaga (Spain). (e-mail: avr@ic.uma.es). Georges Kaddoum is with the École de Technologie  Supérieure, Montreal, Quebec H3C 1K3, Canada, and also with the Cyber Security Systems and Applied AI Research Center, Lebanese American University, Beirut, Lebanon(e-mail: georges.kaddoum@etsmtl.ca).}}

\markboth{Journal of \LaTeX\ Class Files,~Vol.~XX, No.~XX, XX~2024}%
{Shell \MakeLowercase{\textit{et al.}}: Enabling Wideband  Precoding Through 3DCNN}

\maketitle

\begin{abstract}
To meet the ever-increasing demand for higher data rates, 5G and 6G technologies are shifting transceivers to higher carrier frequencies, to support wider bandwidths and more antenna elements. Nevertheless, this solution poses several key challenges: i) increasing the carrier frequency and bandwidth leads to greater channel frequency selectivity in time and frequency domains, and ii) the greater the number of antennas the greater the the pilot overhead for channel estimation and the more prohibitively complex it becomes to determine the optimal precoding matrix. 
This paper presents two deep-learning frameworks to solve these issues. Firstly, we propose a 3D convolutional neural network (CNN) that is based on image super-resolution and captures the correlations between the transmitting and receiving antennas and the frequency domains to combat frequency selectivity. Secondly, we devise a deep learning-based framework to combat the time selectivity of the channel that treats channel aging as a distortion that can be mitigated through deep learning-based image restoration techniques. Simulation results show that combining both frameworks leads to a significant improvement in performance compared to existing techniques with little increase in complexity. 
\end{abstract}


\IEEEpeerreviewmaketitle

\section{Introduction}
\label{sec:intro}

\IEEEPARstart{T}{HE} doubly selectivity of the wireless channel, which manifests in rapid channel fluctuations in time and frequency domain, poses significant challenges to 5G and beyond 5G communication systems, that must perform channel state information (CSI) acquisition to adapt multiple-input multiple-output (MIMO) transmission to channel conditions. 

In 3GPP-compliant 5G networks, the mechanism used for CSI acquisition depends on the duplexing scheme. With the time division duplexing (TDD) scheme considered, the downlink (DL) and uplink (UL) directions share the same frequency band but different time slots. Thus, during a coherence period, the DL CSI information is obtained at the base station (BS) based on the UL channel estimation (CE), which is based on UL sounding reference signals (SRSs), given the channel reciprocity that can be assumed in this scenario. With the frequency division duplexing (FDD) scheme, channel reciprocity cannot be assumed, and thus, the user equipment (UE) transmits a UL CSI report, which is based on the DL CE using CSI reference signals (CSI-RSs) \cite{38.214, Martin2021}. 

In both cases, the frequency selectivity can be combated by increasing the pilot and CSI report overheads in the frequency domain at the expense of having fewer resources available to transmit data. The time selectivity is of particular interest as it leads to channel aging if the time coherence of the channel is less than the minimum periodicity for CSI acquisition that is allowed in the 3GPP standard. For instance, with TDD, the minimum periodicity of the SRS is $2$ ms \cite{38.331}. If the network is using a frequency band at $28$ GHz and the user is moving a $30$ km/h, the time coherence of the channel would be roughly $1.3$ ms, which is smaller than the minimum SRS periodicity that can be configured. 
Techniques that combat channel aging are therefore needed to reduce the pilot periodicity required, and thus the signaling overhead.

\subsection{Related Works}

Artificial intelligence (AI) and machine learning (ML) algorithms, and especially deep learning techniques, have gained a lot of attention in physical (PHY) layer signal processing due to their proven feasibility and superior performance in a wide range of applications.
%
%
In \cite{zhang2017beyond}, Zhang et al. present a residual CNN architecture to deal with the problem of denoising images by learning the noise so it can then be subtracted. 
That concept, which is coined denoising CNN (DnCNN), is also later applied to the PHY layer in \cite{ye2020deep} and \cite{jin2019channel}. In the latter work, the CE task is also taken into account. Another work in which denoising techniques are used in CE is \cite{soltani2019deep}, where a super-resolution NN \cite{dong2015image} \cite{kim2016accurate} is used to map the received pilots to a fully populated representation of the channel before the DnCNN is applied to return a smooth channel matrix. 

Denoising techniques are also applied to compressed sensing in \cite{he2018deep}, where He et al. propose a learned denoising-based approximate message passing (LDAMP) network to perform CE in the millimeter wave (mmWave) range. Takeda et al. \cite{takeda2019mimo} presents a CE approach that considers real-world impairments such as the non-idealities of analog-to-digital converters (ADCs). The use of AI/ML for estimation is not limited to channel matrices but is also suitable for specific channel parameters such as the precoding matrix indicator (PMI) and the channel quality indicator (CQI) \cite{godala2020deep} or the angle of arrival (AoA) and angle of departure (AoD) in MIMO systems \cite{li2020channel}. 

AI/ML techniques have also been utilized to help predict the evolution of a channel over time and its associated fading. The need to predict channel fading is due to the the strong dependence of MIMO systems on an accurate CSI, which can be rendered outdated in rapidly changing environments. Channel fading can be predicted using recurrent NNs (RNNs) \cite{liu2019deep} \cite{madhubabu2019long} or a combination of long short-term memory (LSTM) and a CNN \cite{han2022joint} so that the CSI report can be used to directly predict subsequent fading conditions. In a series of publications \cite{jiang2020deep} \cite{jiang2019recurrent} \cite{jiang2019neural}, Jiang et al. prove that different configurations of RNNs have the potential to predict channel fading with significant improvements in performance over state-of-the-art methods.

A CSI report can contain an enormous amount of information, especially when there is a large number of antennas in the MIMO system. It comes at no surprise that AI/ML tackles the compression task for CSI reporting in a number of ways. Wen et al. \cite{wen2018deep} introduced CSINET, a CNN-based autoencoder that can successfully compress and decompress the CSI by leveraging the naturally occurring structures in the MIMO channel matrices. The decompression stage is the most challenging task, and other mixed architectures \cite{wang2018deep} have been able to outperform CSINET when it comes to decompression by adding RNN to the formula. Since the introduction of the attention mechanism and transformers in 2017 by Vaswani et al. \cite{vaswani2017attention}, compression and feedback have been tackled using either classical transformers \cite{liu2022model} \cite{xu2021transformer} or convolutional-based transformers \cite{bi2022novel}.
This has motivated the industry and standardization bodies such as 3GPP to apply AI/ML to CSI reporting in FDD systems. This approach entails the entire estimated channel matrix being compressed at the UE side prior to transmission, and then reconstructed at the BS side \cite{3gpp_ran112}. 
Although this is a noisy process that adds some error, which is known as channel compression (CC) error, on top of the known CE error, evaluation results show it significantly outperforms classical 5G CSI reporting based on rank indicator (RI) and/or PMI values \cite{3gpp_ran114}. 

Arguably, a critical enabler of achieving large MIMO systems that is also a bottleneck is the beamforming process. The use of higher frequencies for transmission makes beamforming crucial in mmWaves communications and a must in sub-THz communications. However, as the number of antennas in the system increases, it becomes more and more challenging to achieve accurate and precise beamforming due to computation and delay constraints. AI/ML has successfully taken over not only PHY layer functionalities but also beamforming in several ways. The singular value decomposition (SVD) algorithm, which is considered the theoretical upper bound of beamforming, has been replicated with dense NNs \cite{peken2020deep}. Reinforcement learning (RL) has been applied to beamforming in works like \cite{lizarraga2019hybrid} and \cite{wang2018deep}, where analog and digital precoding matrix selection is performed iteratively. 

However, most of the AI/ML-based beamforming approaches that are found in the literature are trained through supervised learning using different architectures, such as NNs for beamforming at the transmitter \cite{huang2019deep}, or both the transmitter and receiver \cite{tao2019constrained}, a one-dimensional (1D) CNN for joint compression sensing and analog beam prediction \cite{li2019deep}, and a two-dimensional (2D) CNNs for analog beamforming \cite{bao2020deep}. Nonetheless, combining CNNs for channel feature extraction and NNs for beam selection from a predefined subset of beams is a compelling approach to beamforming. Elbir et al. have implemented this approach extensively in the context of joint design of precoding and combining matrices \cite{elbir2019cnn}, multiuser scenarios \cite{elbir2019hybrid}, joint precoding and antenna selection \cite{elbir2019joint} and multicarrier precoding \cite{elbir2020low}, all of which are summarized in \cite{elbir2021family}. As is the case for the CSI compression task, the transformer's robust precoding computation and prediction capabilities make it possible to apply the solution proposed in \cite{jiang2022accurate} to more subcarriers than was possible with previous approaches.

\subsection{Contributions}
In this paper, we present a complete AI/ML-based beamforming framework that is resilient to the non-idealities of the CSI report feedback and its adverse effects on the computation of a channel's precoding matrices. The proposed system is successful in a wide range of channel aging conditions and with Gaussian noise that models the CE and CC errors. Our proposed architecture significantly outperforms existing techniques with little increase in complexity. This paper's contributions can be summarized as follows:
\begin{itemize}
    \item We propose a realistic modeling framework for CSI acquisition and precoding in 5G advanced systems. The modeling framework is suitable for TDD and FDD systems, and takes into consideration the effects of channel aging as well as CE error and CC error. 

    \item  We present a novel approach to deal with channel aging. By considering a channel's evolution over time in the correlation time window as noise, it can then be learned and dealt with like any ordinary instance of noise, such as AWGN. We implement an NN-based subsystem that successfully removes Gaussian noise from the received CSI information and compensates for channel aging in the channel's correlation time window. This methodology is radically different than existing works that rely on well-known prediction networks such as RNNs. Instead, we rely on denoising architectures that are less complex but show promising results in terms of BER. 
    
    \item We introduce a three-dimensional (3D) CNN architecture that makes use of the correlation in the frequency domain by adding a dimension to the state-of-the-art CNNs proposed. This architecture can tackle all subcarriers at the same time without scrambling the channel matrices, which makes it fully scalable in a wideband system.    

    \item We demonstrate through simulation results that our proposed AI-based framework that considers channel prediction as an instance of denoising and incorporates 3D CNNs for precoding significantly outperform classical solutions that efficiently combat CE noise, CC noise and  channel aging. To the best of the authors' knowledge, this is the first time that denoising architectures are being used to {mitigate channel aging and 3D CNNs are being applied to these wireless communication problems}. 
\end{itemize}

The rest of this paper is organized as follows. The system model of the 5G system considered is described in Section \ref{sec:System Model}. Then, the proposed 3D CNN for massive MIMO precoding is formulated and described in Section \ref{sec:3D CNN Precoding}. Our proposed framework's performance results are shown and discussed in Section \ref{sec:Results}. Finally, the main conclusions of this work are set out in Section \ref{sec:Conclusions}.

\section{CSI Acquisition and Multi-Antenna Adaptation in the 5G Advanced Framework}
\label{sec:5G advanced}

\begin{figure*}[t!]%
    \centering
    {\includegraphics[width=\textwidth, valign=c]{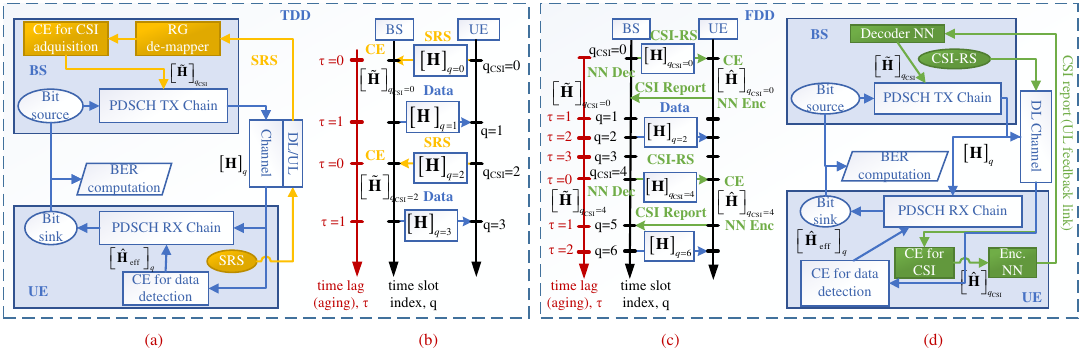} }
    \caption{Block diagrams and time charts related to TDD and FDD CSI acquisition frameworks: (a)  block diagram of 5G data transmission and CSI acquisition at the PHY layer in TDD mode; (b) time chart of CSI acquisition in TDD mode for $T_\mathrm{CSI} = 2$ and $T_\mathrm{offset} = 0$; (c) time chart of CSI acquisition in FDD mode for $T_\mathrm{CSI} = 4$ and $T_\mathrm{offset} = 0$; and (d) block diagram proposed for data transmission and enhanced CSI acquisition at the PHY layer in FDD mode as per the related study item of rel'18 (CSI feedback enhancement) \cite{3gpp_ran114}.}%
    \label{fig:CSI_blocks}%
\end{figure*}
%

In this work, we investigate a 5G advanced link between a BS with $N_t$ transmit antennas and a UE with $N_r$ receive antennas. We focus on the DL direction for data transmission, i.e., the physical downlink shared channel (PDSCH) \cite{Martin2021}, and assume that the number of receive antennas, $N_r$, is less than or equal to the number of transmit antennas, $N_t$, and equal to the number of independent data streams, $N_s$, i.e., $N_s = N_r$ with $N_t \geq N_r$. We consider the 5G resource grid (RG), which divides time into slots that each spans $14$ OFDM symbols. In the frequency domain, the resources are divided into resource blocks (RBs) that each consists of $12$ subcarriers. Thus, the base resource unit, called a resource element (RE), considers a single subcarrier, $k$, of a single OFDM symbol, $\ell$. In 5G, subcarrier spacing (SCS) depends on the numerology and can range from $15$ kHz to $240$ kHz. 
The BS determines the precoding matrix based on the available CSI so as to minimize the BER of a given modulation and coding scheme (MCS). 
We propose a comprehensive 5G advanced CSI acquisition framework to assess the effect of the main impairments, i.e., channel aging, CE error, and CC error. The framework includes a TDD mode and an FDD operation mode and is illustrated in Fig. \ref{fig:CSI_blocks}. 
The block diagram of the transmitter and receiver in both cases is described in Fig. \ref{fig:CSI_blocks}-(a) and Fig. \ref{fig:CSI_blocks}-(d) for TDD mode and FDD mode, respectively. 
%
%
In both cases, the \textit{bit source} block generates a sequence of bits that is transmitted at each time slot, $q$. In 5G, this sequence, which is called a transport block (TB), is received from the medium access control (MAC) layer and coded in accordance with a low-density parity check (LDPC) code with a given coding rate $R$ that enables error correction at the receiver side. Then, the coded bits are mapped into quadrature amplitude modulation (QAM) constellation symbols with $M$ constellation points, i.e., $M$-QAM, and spatially multiplexed using an appropriate precoding matrix. De-modulation reference signals (DMRSs) are also generated and spatially multiplexed so that the effective channel matrix for data detection can be estimated on the receiver. 
Finally, the resulting constellation symbols are mapped to the appropriate REs for OFDM transmission (i.e., RG mapping). The sets $\mathcal{D} \in \mathbb{N}_0^2$ and $\mathcal{P} \in \mathbb{N}_0^2$ identify the REs $(\ell,k)$ that are reserved for data transmission and DMRS transmission, respectively. 
%
%
All these steps are performed by the \textit{PDSCH TX chain} block. 
%
%
On the receiver, i.e., the UE side, the inverse process is performed by the \textit{PDSCH RX chain} block to detect the information bits. The reception process requires performing {CE for data detection} in accordance with the received DMRSs, which is performed by the \textit{CE for data detection} block. At the UE side, the effective channel matrix, which can be expressed as the product of the precoding and channel matrices, is estimated. However, on the BS side, the \textit{CE for CSI acquisition} block instead estimates the actual channel matrix since the CSI-RSs (in FDD mode) and SRSs (in TDD mode) are not multiplied by the precoding matrix. 

The main difference between the FDD and TDD modes pertains to their CSI acquisition processes. In TDD mode, channel reciprocity is assumed to hold, and thus, the UL channel is estimated at the BS side from the SRSs transmitted from the UE. Then, the estimated UL channel matrix is used as the DL channel matrix to determine the optimal precoding matrix for spatial multiplexing, which involves a different beam being used to transmit each layer (i.e., stream).

To reduce the CSI acquisition overhead in the UL direction, the SRSs that are transmitted periodically in a given slot $q$ satisfy the following equation \cite{38.211} 
\begin{equation}
    \label{eq:SRS_lots}
    (q - T_\mathrm{offset}) \mod T_\mathrm{CSI} = 0,
\end{equation}

\noindent where $T_\mathrm{CSI}$ stands for the SRS periodicity, and $T_\mathrm{offset}$, for the slot offset. Therefore, for data transmission at slot $q$, the BS uses a precoding matrix that is based on the channel matrix that was estimated from the last SRS received in slot $q_\mathrm{CSI}$. This latter slot can be expressed as
\begin{equation}
    \label{eq:q_CSI}
    q_\mathrm{CSI} = f_\mathrm{CSI}(q) = \floor*{\frac{q - T_\mathrm{offset}}{T_\mathrm{CSI}}} T_\mathrm{CSI} + T_\mathrm{offset}.
\end{equation}

The time lag between the current slot, $q$, (i.e., the slot in which data transmission takes place) and the previous CSI slot, $q_\mathrm{CSI}=f_\mathrm{CSI}(q)$, is referred to as \textit{aging} and is expressed as $\tau = q - f_\mathrm{CSI}(q)$. 
Fig. \ref{fig:CSI_blocks}-(c) illustrates the first four slots of a time chart for CSI acquisition and data transmission in TDD mode with $T_\mathrm{CSI}=2$ and $T_\mathrm{offset}=0$. This means that the SRSs are being transmitted in slots $q_\mathrm{CSI} = \{0, 2,...\}$. Assuming channel reciprocity holds, the equivalent DL channel matrix, $\left[ \mathbf{\tilde{H}} \right]_{q_\mathrm{CSI}} \in \mathbb{C}^{N_{\rm symb}^{\rm slot} \times N_\mathrm{FFT} \times N_r \times N_t}$, is estimated at the BS side in those slots $q_\mathrm{CSI}$ that fulfill \eqref{eq:q_CSI}. Therefore, $\mathbf{\tilde{H}} \in \mathbb{C}^{N_\mathrm{slot} \times N_{\rm symb}^{\rm slot} \times N_\mathrm{FFT} \times N_r \times N_t}$, stands for a tensor that stacks the equivalent DL channel matrix that is available at every slot in a simulation of $N_\mathrm{slot}$ slots, where $N_{\rm symb}^{\rm slot}$, and $N_\mathrm{FFT}$ refers to the number of symbols per slot, and the fast Fourier transform (FFT) size, respectively. The estimated channel matrix, $\left[ \mathbf{\tilde{H}} \right]_{q_\mathrm{CSI}}$, is subject to estimation errors, $\left[ \mathbf{Z^\mathrm{UL,CE}} \right]_{q_\mathrm{CSI}}$, and can thus be expressed as
\begin{equation}
    \label{eq:TDD_estimated_CSI}
    \left[ \mathbf{\tilde{H}} \right]_{q_\mathrm{CSI}} = \left[ \mathbf{{H}} \right]_{q_\mathrm{CSI}} + \left[ \mathbf{Z^\mathrm{UL,CE}} \right]_{q_\mathrm{CSI}},
\end{equation}

\noindent where $\left[ \mathbf{{H}} \right]_{q_\mathrm{CSI}}$ is the ideal channel frequency response (CFR) of the channel at slot $q_\mathrm{CSI}$. The estimated channel matrix represents the available CSI information for determining the optimal precoding matrix and is thus used in subsequent data transmissions until a new channel estimate is obtained in the next CSI slot.  

The actual channel's evolution in subsequent slots involves a decrease in performance since the actual channel, $\left[ \mathbf{{H}} \right]_{q}$, differed from the estimated channel in the previous slot. This error due to channel aging can be expressed as 

\begin{equation}
    \label{eq:Z^agging}
    \left[ \mathbf{{Z}}^\mathrm{aging} \right]_{q} = \left[ \mathbf{{H}} \right]_{q} - \left[ \mathbf{{H}} \right]_{q_\mathrm{CSI}},
\end{equation}

\noindent where $q_\mathrm{CSI} = f_\mathrm{CSI}(q)$ as per \eqref{eq:q_CSI}. 

In FDD mode, channel reciprocity does not hold, and hence, CSI acquisition requires that channel estimation be completed at the UE side. 
Fig. \ref{fig:CSI_blocks}-(d) shows the block diagram of CSI acquisition in FDD mode, with the BS transmitting the CSI-RSs to estimate the channel for CSI acquisition. Then, the estimated channel matrix, $\left[ \mathbf{\hat{H}} \right]_{q_\mathrm{CSI}}$, is compressed and transmitted to the BS via a return channel. The BS has to decompress the estimated channel matrix, which is a noisy process that adds errors. This framework, which differs from the classical 5G framework that relies on CQI, PMI and RI reports, was initially proposed and investigated by the 3GPP as part of a study item called \textit{AI/ML for NR Air Interface} for release 18 \cite{3gpp_ran112, 3gpp_ran114}. In such a study item, the compression and reconstruction were achieved with autoencoders that included encoding and decoding NNs \cite{wen2018deep}. 

Therefore, in FDD mode, the available channel matrix that is needed to determine the optimal precoding for data transmission in slot $q$ can be expressed as
\begin{equation}
    \label{eq:CE_CC_H}
    \left[ \mathbf{\tilde{H}} \right]_{q_\mathrm{CSI}} = \left[ \mathbf{{H}} \right]_{q_\mathrm{CSI}} + \left[ \mathbf{Z^\mathrm{DL,CE}} \right]_{q_\mathrm{CSI}} + \left[ \mathbf{Z^\mathrm{CC}} \right]_{q_\mathrm{CSI}},
\end{equation}

\noindent where $\left[ \mathbf{Z^\mathrm{CC}} \right]_{q_\mathrm{CSI}}$ is the CC error and $\left[ \mathbf{Z^\mathrm{DL,CE}} \right]_{q_\mathrm{CSI}}$ is the CE error determined from the DL CSI-RSs and $q_\mathrm{CSI} = f_\mathrm{CSI}(q)$. Finally, the time chart shown in Fig. \ref{fig:CSI_blocks}-(c), which represents the first seven slots of a link configured with $T_\mathrm{CSI} = 4$ and $T_\mathrm{offset} = 0$, illustrates the dynamics of the CSI acquisition process in FDD mode. It is observed that a longer CSI period, $T_\mathrm{CSI}$, is associated with higher channel aging values, $\tau$. In addition, the slot before the CSI slot, i.e., $\tau=T_\mathrm{CSI}-1$, is when data transmission is affected more by the decrease in performance due to channel aging. 

\section{System Model}
\label{sec:System Model}

\begin{figure}[t!] 
\centering
  \includegraphics[width=0.8\columnwidth ]{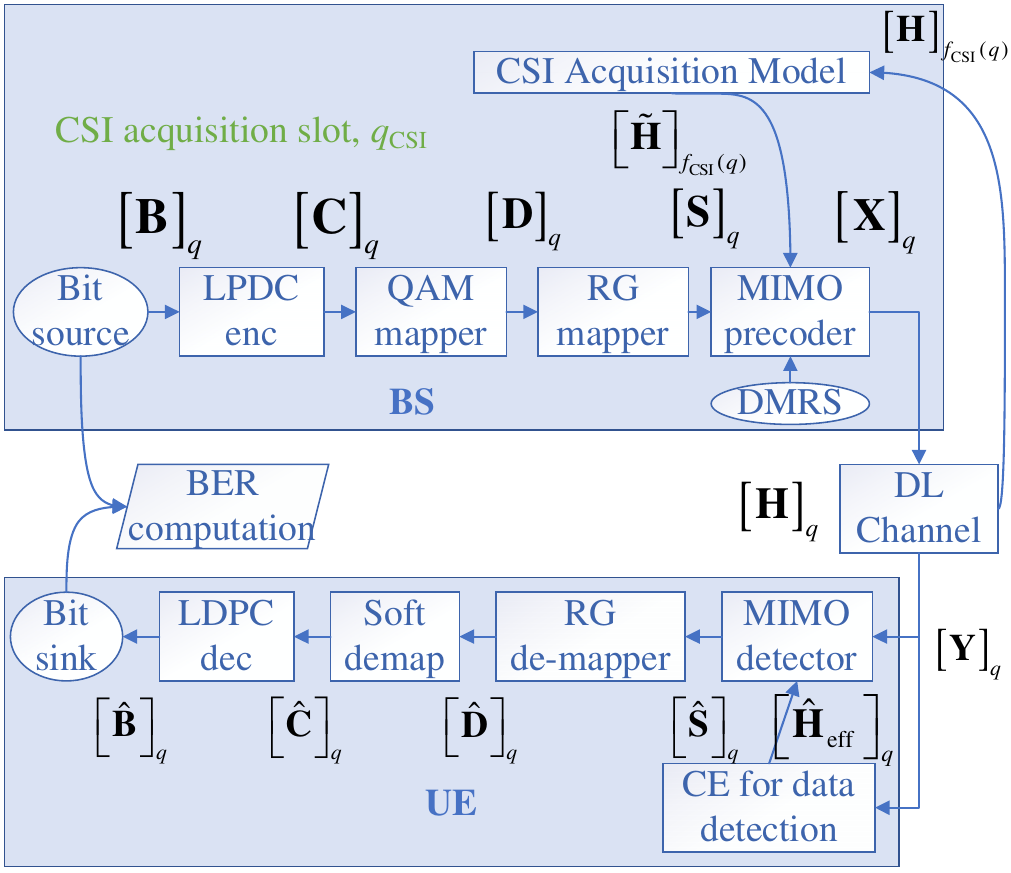}
  \caption{Block diagram of the implemented model (abstracted system model).}
  \label{fig:system}
\end{figure}

In this section, we present a system model that combines the previous TDD and FDD CSI acquisition procedures in a single modeling framework. This model is illustrated in Fig. \ref{fig:system}. As mentioned in the previous section, the CSI information that is available at the BS side is degraded by two types of impairment: i) CSI acquisition error, and ii) CSI aging. In TDD mode, error stems mainly from CE being  based on the UL SRSs at the BS side as per \eqref{eq:TDD_estimated_CSI}. In FDD mode, on the other hand, the error is due to CE being based on the DL CSI-RSs and to CC, as per \eqref{eq:CE_CC_H}. In both cases, the error can be modeled as an additive term, $\left[ \mathbf{Z}^\mathrm{CSI} \right]_{q_\mathrm{CSI}}$, with a given distribution. 
We assume the CSI acquisition error has a normal distribution. Therefore, the CSI acquisition error of the available channel matrix $\left[ \mathbf{\tilde{H}} \right]_q$ for the $q$-th slot, $\ell$-th OFDM symbol, $k$-th subcarrier, $n_r$-th receive antenna and $n_t$-th transmit antenna is distributed as
$\left[ \mathbf{Z}^\mathrm{CSI} \right]_{q, \ell, k, n_r, n_t} \sim \mathcal{CN} (0, \sigma_\mathrm{CSI})$, where $\sigma^2_\mathrm{CSI}$ is the error power. 
As for the channel aging, it is due to the precoding matrix at slot $q$ being based on the CSI acquired  at slot $q_\mathrm{CSI} = f_\mathrm{CSI}(q)$. Therefore, the CSI available at slot $q$ can be expressed as
\begin{equation}
    \label{eq:CSI_acq_model}
    \left[ \mathbf{\tilde{H}} \right]_{q} = \left[ \mathbf{{H}} \right]_{f_\mathrm{CSI}(q)} + \left[ \mathbf{Z^\mathrm{CSI}} \right]_{f_\mathrm{CSI}(q)},
\end{equation}
\noindent which is modeled by the \textit{CSI acquisition model} block in Fig. \ref{fig:system}. 

The frequency domain channel matrix is normalized so that $\left[ \mathbf{H}\right]_{q, \ell, k, n_r, n_t} \sim \mathcal{CN} (0, 1)$ and $\mathbb{E} \left[ \left\lVert  \left[ \mathbf{{H}} \right]_{q,\ell,k}  \right\rVert^2 _F \right] = N_t N_r$. The signal-to-noise ratio (SNR) of CSI acquisition is defined as the quotient between the power of the ideal channel at a given RE ($\ell, k$) and the antenna pair ($n_r, n_t$) and expressed as 
\begin{equation}
\label{eq:SNR_CSI}
    \mathrm{SNR}^\mathrm{CSI}=\frac{\mathbb{E} \left[\left\lvert \left[ \mathbf{H}\right]_{q, \ell, k, n_r, n_t} \right\rvert^2 \right]}{\mathbb{E} \left[\left\lvert \left[ \mathbf{Z}^\mathrm{CSI}\right]_{q, \ell, k, n_r, n_t} \right\rvert^2 \right]}=\sigma^{-2}_\mathrm{CSI}.
\end{equation}

CSI acquisition is necessary for the \textit{PDSCH TX chain} block shown in Fig. \ref{fig:CSI_blocks} to perform multi-antenna adaptation for the data transmission process. Fig. \ref{fig:system} illustrates the sub-blocks that form the aforementioned \textit{PDSCH TX chain} and \textit{PDSCH RX chain} blocks. 
These blocks were developed using Sionna design principles, and thus, each block processes a set of $N_\mathrm{slots}$ slots (i.e., the batch size), which is the first dimension of each tensor. The aim of each of those blocks' sub-blocks is summarized as follows:


\begin{itemize}
    \item \textit{Bit source}: This sub-block generates a sequence of bits that is transmitted at each time slot $q$. In 5G, this sequence, which is known as the TB, is received from the MAC layer and expressed as $\left[\mathbf{B} \right]_q \in \{0, 1\}^{N_s \times K_{\rm CBS}}$, where $N_s$ is the number of layers (i.e., independent streams) and $K_\mathrm{CBS}$ is the code block size (CBS)\footnote{5G specifications states the code block segmentation process for TBs, which depends on the TB size and coding rate as per \cite{38.212}, Section 5.2.2. For the sake of modeling simplicity, our implementation assumes that the TB is segmented into $N_s$ code blocks.}. Therefore, a full tensor, $\mathbf{B}\in \{0, 1\}^{N_\mathrm{slot} \times N_s \times K_{\rm CBS}}$, is generated by this sub-block for either simulation (i.e., testing) or training. 
    
    \item \textit{LDPC encoder}: This sub-block encodes the TB bits in accordance with an LDPC code, which enables error correction at the receiver side. Thia sub-block comprises LDPC base graph selection, LDPC coding, and rate matching to achieve the target code rate $R$, which is determined by the MCS \cite{38.212}. The code rate is expressed as $R=K_{\rm CBS}/N_{\rm CB}$, where $N_{\rm CB}$ represents the number of bits that are effectively coded after LDPC coding and rate matching. The coded block is written as $\mathbf{C} \in \{0, 1\}^{N_\mathrm{slot} \times N_s \times N_{\rm CB}}$. 

    \item \textit{QAM mapper}: This sub-block maps the sequence of coded bits to the selected $M$-QAM constellation symbols, where $M \in \{4,16,64,256\}$. The transmitted sequence of constellation symbols in a simulation of $N_\mathrm{slot}$ slots is expressed in matrix form as $\mathbf{D} \in \mathcal{C}^{N_\mathrm{slot} \times N_s \times \lvert \mathcal{D} \lvert}$, where $\mathcal{C}$ represents the constellation symbols and \mbox{$\mathcal{D}$ stands} for the set of REs allocated to data transmission. The constellation symbols have unit power, and thus, $\mathbb{E} \left[ \lvert s \lvert^2  \right] = 1, \forall s \in \mathcal{C}$. 

    \item \textit{RG mapper}: At time slot $q$, this sub-block maps the sequence of constellation symbols for every layer, $\left[ \mathbf{C} \right]_q$, to the REs reserved for data transmission, i.e., $(\ell, k) \in \mathcal{D}$. This sub-block also generates the DMRSs and maps them to the REs reserved for DMRS transmission, \mbox{i.e.,  $(\ell, k) \in \mathcal{P}$}. The output tensor is written as $\mathbf{S} \in \mathbb{C}^{N_\mathrm{slot} \times N^\mathrm{slot}_\mathrm{symb} \times N_\mathrm{FFT} \times N_s}$, where  $N^\mathrm{slot}_\mathrm{symb}$ and $N_\mathrm{FFT}$ are the number of OFDM symbols per slot and the FFT size, respectively.
    
    \item \textit{MIMO precoder}: This sub-block receives one input tensor of constellation symbols per layer and RE, and performs spatial multiplexing, which involves multiplying each layer by an appropriate beamforming vector that is determined from the acquired CSI information. The output tensor can be written as $\mathbf{X} \in \mathbb{C}^{N_\mathrm{slot} \times N^\mathrm{slot}_\mathrm{symb} \times N_\mathrm{FFT} \times N_t}$. The computation of the beamforming vectors that form the precoding matrix is explained in detail later in this section. 
    
    \item \textit{DL channel}: This sub-block models the multipath channel and noise generation. The multipath channel follows the 3GPP NR cluster delay line (CDL) model as per \cite{38.901}. The channel's ideal CFR is expressed as $\mathbf{H} \in \mathbb{C}^{N_\mathrm{slot} \times N^\mathrm{slot}_\mathrm{symb} \times N_\mathrm{FFT} \times N_r \times N_t}$. Once the transmitted signal has passed through the multipath channel, AWGN is added to achieve an average DL SNR for each RE that is reserved for data or DMRS transmission and for each antenna pair ($n_t, n_t$) that can be expressed as $\mathrm{SNR}=1/\sigma_n^2$, where $\sigma_n^2$ is the noise power, since $\mathbb{E} \left[\left\lvert \left[ \mathbf{H}\right]_{q, \ell, k, n_r, n_t} \right\rvert^2 \right] = 1$. The output tensor has the following shape: $\mathbf{Y} \in \mathbb{C}^{N_\mathrm{slot} \times N^\mathrm{slot}_\mathrm{symb} \times N_\mathrm{FFT} \times N_r}$.
    
    \item \textit{CE for data detection}: This sub-block performs CE based on the transmitted DMRSs. Thus, it estimates the effective channel matrix, which takes into account both the CFR and the precoding matrix. The output tensor is the estimated effective channel and is written as $\mathbf{\hat{H}}_\mathrm{eff} \in \mathbb{C}^{N_\mathrm{slot} \times N^\mathrm{slot}_\mathrm{symb} \times N_\mathrm{FFT} \times N_r \times N_s}$.

    \item \textit{MIMO detector}: This sub-block uses $\mathbf{\hat{H}}_\mathrm{eff}$ to estimate the transmitted constellation symbols for each layer. Therefore the output tensor, $\mathbf{\hat{S}}$ has the same shape as $\mathbf{S}$.

    \item \textit{RG demapper}: This sub-block performs the inverse process of the \textit{RG mapper}, and thus outputs a tensor $\mathbf{\hat{D}}$ of equalized constellation symbols with the same shape as  $\mathbf{{D}}$.

    \item \textit{Soft demapper}: This sub-block estimates the log-likelihood ratio (LLR) of the equalized constellation symbols from the effective noise power estimated after MIMO detection. The LLR value includes a hard decision for each coded bit, with the sign, and a measure of the reliability of each decision with its absolute value. Therefore, this sub-block's output tensor can be written as $\mathbf{\hat{C}} \in \mathbb{R}^{N_\mathrm{slot} \times N_s \times N_{\rm CB}}$.

    \item \textit{LDPC decoder}: This sub-block iteratively decodes the LDPC code using the flooding version of the sum-product algorithm (SPA) \cite{2004Ryan}, where all nodes are updated in a parallel fashion. It also uses the boxplus-phi function as a check node update rule. The tensor of decoded information bits, $\mathbf{\hat{B}}$, has the same shape as $\mathbf{B}$. These two tensors are compared to obtain the BER of a given simulation. 
\end{itemize}

\subsection{MIMO Precoding}
\label{sec:precoder}
Spatial diversity in MIMO systems is achieved through spatial multiplexing, in which a set of $N_s$ independent data streams are transmitted over different beams. This process requires CSI acquisition to be performed as discussed in the previous section in order to adapt the beams to varying instantaneous channel conditions. 
In this paper, we use the term \say{precoder} and the variable, $ \mathcal{F}(\cdot)$, to denote the signal processing block that computes the appropriate precoding matrix for slot $q$ at the RE $(\ell, k)$ based on the CSI available in that slot, $\left[\mathbf{\tilde{H}} \right]_q$,

\begin{equation}
    \label{eq:precoding}
    \left[\mathbf{{F}} \right]_{q,\ell, k} = \mathcal{F} \left(\left[\mathbf{\tilde{H}} \right]_{q,\ell, k} \right),
\end{equation}

\noindent where $ \left[\mathbf{{F}} \right]_{q,\ell, k} \in \mathbb{C}^{N_t \times N_s}$ is the precoding matrix that is computed from the available CSI,  $\left[\mathbf{\tilde{H}} \right]_{q,\ell, k} \in \mathbb{C}^{N_r \times N_t}$, with $\left[\mathbf{\tilde{H}} \right]_{q}$, being expressed in \eqref{eq:CSI_acq_model}. This precoding matrix is applied by the \textit{MIMO precoder} block (see Fig. \ref{fig:system}) to each RE to wich either data or a DMRS is mapped. Therefore, the  constellation symbols transmitted on those REs, $(\ell,k) \in \mathcal{D} \cup \mathcal{P}$, can be expressed as

\begin{align}
    \label{eq:X}
    \left[\mathbf{X} \right]_{q,\ell, k} &=  \left[\mathbf{{F}} \right]_{q,\ell, k} \left(  \left[\mathbf{{S}} \right]_{q,\ell, k} \right)^T \\ \nonumber
    &= \sum_{i=0}^{N_s-1}  \left[\mathbf{{F}} \right]_{q,\ell, k, i, :} \left[\mathbf{{S}} \right]_{q,\ell, k, i},
\end{align}

\noindent where $ \left[\mathbf{{S}} \right]_{q,\ell, k, i} \in \mathcal{C} \subset \mathbb{C}$ represents the scalar constellation symbol being transmitted in the $(i+1)$-th layer from the RE $(\ell,k)$ of the $q$-th slot, whereas $\left[\mathbf{{F}} \right]_{q,\ell, k, i, :} \in \mathbb{C}^{N_t}$ represents the beamforming vector of the $(i+1)$-th layer. 

We have considered the practical regularized zero-forcing (ZF) precoding \cite{2017Hoydis} as our baseline, which uses regularization to improve the numerical stability of the pseudo-inverse. It also ensures that each stream is precoded with a unit-norm vector, which makes it possible to respect the maximum transmit power constraints imposed by practical transmitters. 
\begin{align}
    \label{eq:ZF_precoder}
    \mathcal{F}^{\mathrm{ZF}} \left(\left[\mathbf{\tilde{H}} \right]_{q,\ell, k} \right) &= \left[\mathbf{V} \right]_{q,\ell, k} \left[\mathbf{ \Sigma} \right]_{q,\ell, k}\\ \nonumber
    \left[\mathbf{V} \right]_{q,\ell, k} &=  \left[\mathbf{\tilde{H}} \right]_{q,\ell, k}^{H}(\left[\mathbf{\tilde{H}} \right]_{q,\ell, k} \left[\mathbf{\tilde{H}} \right]_{q,\ell, k}^{H})^{-1} \\ \nonumber
    \left[\mathbf{ \Sigma} \right]_{q,\ell, k} &= \mathrm{diag} \left(\left\lVert \left[\mathbf{V} \right]_{q,\ell, k,i} \right\lVert_2^{-1}, i=0,..,N_s-1 \right)  \nonumber
\end{align}

The precoding method's performance is highly dependent on the accuracy of the available CSI information, $\left[\mathbf{\tilde{H}} \right]_{q,\ell, k}$. As the time lag (aging) between the current slot $q$ and the last CSI slot increases, $\tau = q - f_\mathrm{CSI}(q)$, the performance of spatial multiplexing degrades considerably. There are several implementation issues, such as the processing time of CSI reports in FDD mode, or the fact that the SRSs are transmitted in UL slots but the estimated channel is used in DL slots in TDD mode, which makes $\tau > 0$ and thus greatly reduces the performance of spatial multiplexing. 

\subsection{Channel Estimation and MIMO Detection} 
\label{sec:CE_data}

The constellation symbols received by the UE antenna array at slot $q$ and RE $(\ell, k)$ can be expressed in matrix form as
\begin{align}
    \label{eq:Y}
    \left[\mathbf{Y} \right]_{q,\ell, k} &= \left[\mathbf{H} \right]_{q,\ell, k} \left[\mathbf{X} \right]_{q,\ell, k} 
        + \left[\mathbf{Z}^\mathrm{DL} \right]_{q,\ell, k}  \\
    &= \underbrace{\left(  \left[\mathbf{H} \right]_{q,\ell, k} \left[\mathbf{F} \right]_{q,\ell, k} \right)}_{\left[\mathbf{H}_\mathrm{eff} \right]_{q,\ell, k}}
            \left(  \left[\mathbf{S} \right]_{q,\ell, k} \right)^T + \left[\mathbf{Z}^\mathrm{DL} \right]_{q,\ell, k},  \nonumber
\end{align}

\noindent where $\left[\mathbf{Z}^\mathrm{DL} \right]_{q,\ell, k} \sim \mathcal{CN}(0, \sigma_n^2)$ represents AWGN noise, $\left[\mathbf{H} \right]_{q,\ell, k} \in \mathbb{C}^{N_r \times N_t}$ is the channel matrix in the frequency domain, $\left[\mathbf{F} \right]_{q,\ell, k} \in \mathbb{C}^{N_t \times N_s}$ is the precoding matrix, and $\left[\mathbf{H}_\mathrm{eff} \right]_{q,\ell, k} \in \mathbb{C}^{N_r \times N_s}$ is the effective channel matrix. 
%

To perform data detection, the receiver needs to estimate the effective channel matrix, $\left[\mathbf{H}_\mathrm{eff} \right]_{q,\ell, k}$. This requires a set of $N_s$ orthogonal pilots to be transmitted at each coherence resource group (CRG) for each OFDM symbol with pilots. A CRG is a set of contiguous REs whose channel is assumed to be constant. In this work, we assume a \textit{Kronecker} pilot pattern, which ensures orthogonality among the transmitted pilots in the frequency domain. Therefore, the $g$-th CRG $\mathcal{R}_{g}$ involves a set of $N_s$ consecutive subcarriers. With the Kronecker pilot pattern, the vector $\mathbf{L} = (\ell_0, .., \ell_{\lvert \mathbf{L} \lvert - 1})$ indicates which OFDM symbols are reserved for DMRS transmission in time domain, whereas in the frequency domain, all the active subcarriers, $N_A$, of those OFDM symbols are reserved for DMRS transmission. The number of active subcarriers is expressed as $N_A = N_\mathrm{FFT} - \lvert  \mathbf{k}_\mathrm{guard}\lvert$, where $\mathbf{k}_\mathrm{guard} = \left[k_{-}, k_{+} \right]$ is the vector that defines the number of guard band subcarriers below $k_{-}$ and above $k_{+}$ the direct current (DC) frequency. Therefore, with the Kronecker pilot pattern, the REs reserved for DMRS transmission and the REs of the $g$-th CRG are written as 
\begin{align}
    \label{eq:Kronecker_REs}
    \mathcal{R}_{g} &= \left\{ (\ell, k) \in \mathbb{N}_0 : \ell \in \mathbf{L}, k \in \left[g N_s, \left(g + 1\right) N_s - 1 \right] \subset \mathbb{N}_0 \right\}, \nonumber \\
    &\; \mathrm{with} \; g \in \mathcal{G} = \left[0, \floor*{\frac{N_A}{N_s}} - 1 \right] \subset \mathbb{N}_0 ,  \\
    \mathcal{P} &= \bigcup_{g \in \mathcal{G}}\mathcal{R}_{g} 
        = \left\{ (\ell, k) \in \mathbb{N}_0 : \ell \in \mathbf{L}, k \in \left[0, N_A - 1 \right] \subset \mathbb{N}_0 \right\}. \nonumber
\end{align}

\begin{figure}[t!]%
    \centering
    \subfloat[]{{\includegraphics[width=0.4\columnwidth, valign=c]{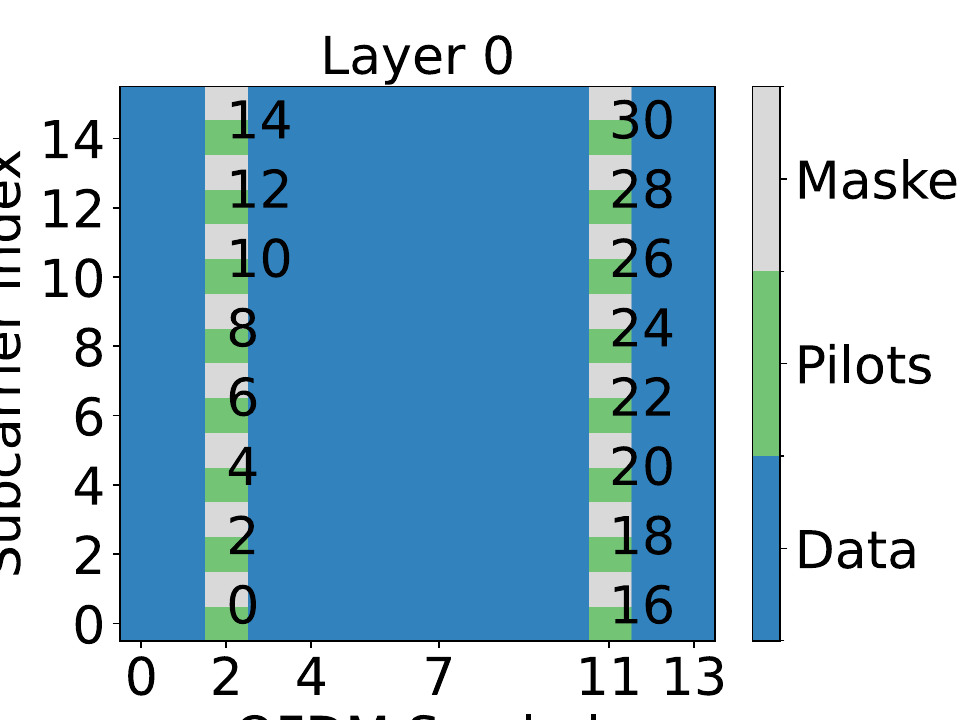} }}%
    \subfloat[]{{\includegraphics[width=0.4\columnwidth, valign=c]{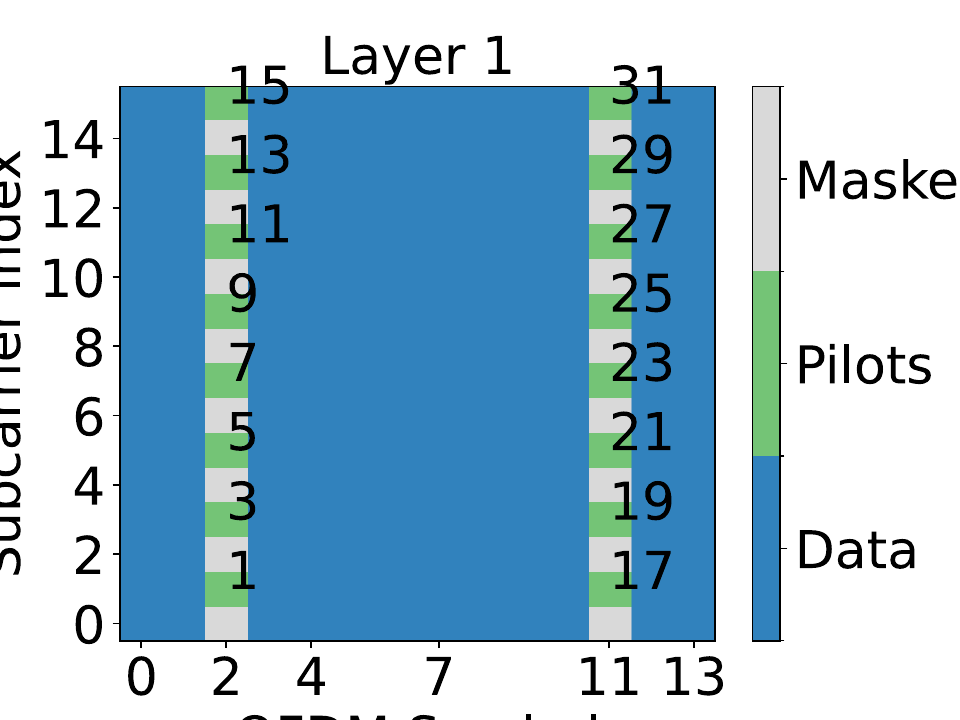} }}%
    \caption{Kronecker pilot pattern for $N_\mathrm{FFT} = 16, \mathbf{k}_\mathrm{guard}=[0,0], N_s=2$, and $\mathbf{L}=(2, 11)$. A zero power, i.e., $0$,  symbol is transmitted on the masked REs.}%
    \label{fig:pilot_pattern}%
\end{figure}

If the $(\ell+1)$-th OFDM symbol contains a DMRS, the constellation symbols transmitted in the $N_s$ layers, 
$\left[\mathbf{S} \right]_{q,\ell, k} = \left(s_0, .., s_{N_s-1} \right)$, with $s_\nu = \left[\mathbf{S} \right]_{q,\ell, k, \nu}$, can be written as
\begin{equation}
\label{eq:pilot_seq}
    \left[\mathbf{S} \right]_{q,\ell, k, \nu} =
    \begin{cases}
        p_n & \text{if } \, k \, \mathrm{mod} \, N_s = \nu \\
        0 & \text{otherwise },
    \end{cases}
\end{equation}

\noindent where $s_\nu$ is the constellation symbol transmitted in the \mbox{$(\nu+1)$-th} layer and $p_n$ is a QPSK-modulated known sequence. Thus, according to \eqref{eq:pilot_seq}, the vector of transmitted constellation symbols in OFDM symbols with DMRSs can be written as $\left[\mathbf{S} \right]_{q,\ell, k, \nu} = \left(0, .., 0, p_n, 0, .., 0 \right)$, with the $(\nu+1)$-th element being $p_n$ with $\nu= k \, \mathrm{mod} \, N_s$ and the rest of the elements being $0$. The sequence index, $n$, of an OFDM symbol with a DMRS, $\ell \in \mathbf{L}$ can be written as 
\begin{align}
    \label{eq:sequence_index_n}
    n &= f_\mathrm{seq}(k,\ell, \mathbf{L}) = k + N_A f_\mathrm{index}\left( \ell, \mathbf{L}\right), \nonumber \\
    i &= f_\mathrm{index}\left( \ell_i, (\ell_0, .., \ell_i, .., \ell_{\lvert  \mathbf{L} \lvert}) )\right),
\end{align}

\noindent where $f_\mathrm{index}\left( \ell, \mathbf{L}\right)$ returns the index $i$ if $\ell \in \mathbf{L}$ is the $(i+1)$-th element of $\mathbf{L}$. Fig. \ref{fig:pilot_pattern} illustrates the Kronecker pilot pattern for $N_A = N_\mathrm{FFT} = 16, N_s=2$, and $\mathbf{L}=(2, 11)$.

A least squares (LS) CE is performed to estimate the $(n_r+1,\nu+1)$-th element of the effective channel matrix $\left[\mathbf{H}_\mathrm{eff} \right]_{q,\ell, k} \in \mathbb{C}^{N_r \times N_s}$ as follows \cite{2017Hoydis}
\begin{equation}
    \label{eq:LS_CE}
    \left[\mathbf{\hat{H}}_\mathrm{eff} \right]_{q,\ell, k, n_r, \nu} = \frac{\left[\mathbf{Y} \right]_{q,\ell, k, n_r}}{p_n} = \frac{\left[\mathbf{Y} \right]_{q,\ell, k, n_r}}{\lvert p_n \lvert^2} p_n^*,
\end{equation}

\noindent where $n = f_\mathrm{seq}(k,\ell, \mathbf{L})$ and $\nu = k \mod N_s$. Hence, the transmitted vector $\left(\left[\mathbf{S} \right]_{q,\ell, k}\right)^t \in \mathbb{C}^{N_s \times 1}$ makes it possible to estimate the $(\nu+1)$ column of the effective channel matrix $\left[\mathbf{H}_\mathrm{eff} \right]_{q,\ell, k} \in \mathbb{C}^{N_r \times N_s}$. Therefore, a single estimation of the matrix $\left[\mathbf{H}_\mathrm{eff} \right]_{q,\ell, k}$ is obtained for each CRG. A nearest neighbor interpolation is performed in the time domain to obtain the estimated effective channel matrix for the REs reserved for data transmission. 

The equalization follows a linear minimum mean squared error (LMMSE) criteria \cite{Yi2011}, and thus, the constellation data symbols detected at slot $q$ and RE $(\ell, k)$ are obtained as 
\begin{align}
    \label{eq:equalized_signal}
    & \left[\mathbf{\hat{X}} \right]_{q,\ell, k} = \left[\mathbf{Y} \right]_{q,\ell, k} \left[\mathbf{W}_\mathrm{mmse} \right]_{q,\ell, k} \\
    & \left[\mathbf{W}_\mathrm{mmse}\right]_{q,\ell, k} = \mathrm{diag}\left(\left[\mathbf{G} \right]_{q,\ell, k} \left[\mathbf{\hat{H}}_\mathrm{eff} \right]_{q,\ell, k} \right)^{-1} \left[\mathbf{G} \right]_{q,\ell, k} \nonumber \\
    & \left[\mathbf{G} \right]_{q,\ell, k} = \left[\mathbf{\hat{H}}_\mathrm{eff} \right]_{q,\ell, k}^H \left(\left[\mathbf{\hat{H}}_\mathrm{eff} \right]_{q,\ell, k}^H \left[\mathbf{\hat{H}}_\mathrm{eff} \right]_{q,\ell, k} + \left[\mathbf{\hat{\Sigma}}_n^2\right]_{q,\ell, k} \right)^{-1},  \nonumber
\end{align}

\noindent where $\left[\mathbf{\hat{\Sigma}}_n^2\right]_{q,\ell, k} = \mathbb{E}\left[ \mathbf{n} \mathbf{n}^H \right]$ is the estimated noise correlation matrix, with 
$\mathbf{n} = \left[\mathbf{Z}^\mathrm{DL}\right]_{q,\ell, k}$, and $\left[\mathbf{W}_\mathrm{mmse}\right]_{q,\ell, k} \in \mathbb{C}^{N_s \times N_r}$ is the detection matrix. 

\section{Proposed Neural Network Solution}
\label{sec:3D CNN Precoding}
\begin{figure}[h!]
  \centering
  \includegraphics[width=0.6\columnwidth ]{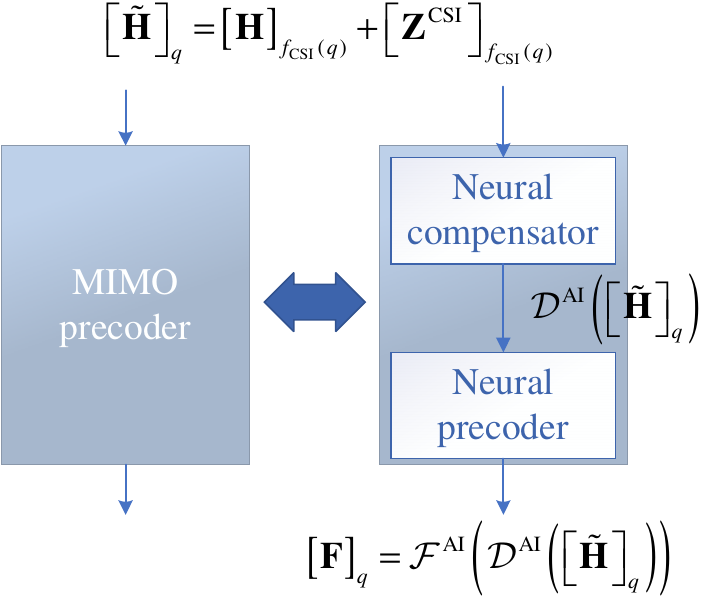}
  \caption{Subsystems in the proposed solution.}
  \label{fig:precoder_insigth}
\end{figure}

In this section, we explain in detail our proposed AI and deep NN. The solution is comprised of two subsystems, namely the neural compensator and neural precoder, which are both part of the \textit{MIMO precoder} block shown in Fig. \ref{fig:system} and are connected as shown in Fig. \ref{fig:precoder_insigth}.

The aim of the neural compensator is to mitigate the main impairments that affect the available CSI information, i.e., aging, CE error and CC error, and that of the neural precoder is to efficiently compute the precoding matrices given an input channel. These two operations are linked as follows:
\begin{equation}
    \label{eq:operation_compensator}
    \left[ {\bf{F}} \right]_{q}  = {\cal F}^{{\rm{AI}}} \left( \left[ \mathbf{\check{H}} \right]_{q} \right) , \quad 
    \left[ \mathbf{\check{H}} \right]_{q} = {\cal D}^\mathrm{AI}\left(\left[ \mathbf{\tilde{H}} \right]_{q} \right),
\end{equation}

\noindent where $\left[ \mathbf{\check{H}} \right]_{q} = {\cal D}^\mathrm{AI}\left(\left[ \mathbf{\tilde{H}} \right]_{q} \right) \in \mathbb{C}^{N_\mathrm{symb}^\mathrm{slot} \times N_\mathrm{FFT} \times N_r \times N_t}$ refers to the compensator's denoising operation, and ${\cal F}^{{\rm{AI}}} \left( \left[ \mathbf{\ddot{H}} \right]_{q}\right) \in \mathbb{C}^{N_\mathrm{symb}^\mathrm{slot} \times N_\mathrm{FFT} \times N_t \times N_s}$ represents the neural precoding. In an abuse of notation, we refer here to the precoding operation for all the symbols and subcarriers of slot $q$. 

Thus, the channel being used to compute the precoding matrices is an estimated version of the current channel, rather than the impaired version that is available to baseline precoders. Note that both operations account for the whole frequency domain, which eliminates the need to recompute the precoding matrices on a per subcarrier basis or to reuse the precoding matrices of one subcarrier across the whole set.

\subsection{Neural Compensator}



\subsubsection{Problem Formulation}

We aim to design the neural compensator, $\mathcal{D}^\mathrm{AI} \left(\left[ \mathbf{\tilde{H}} \right]_{q} ; \mathbf{\Theta}_{\mathcal{D}} \right)$ and to find the set of parameters $\mathbf{\Theta}_{\mathcal{D}}$ that minimizes the mean square error (MSE) between the compensated channel at slot $q$ (which suffers from aging and CE error) and the ground truth, i.e., the actual channel, such that

\begin{equation}
    \label{eq:goal_compensator}
    \arg\min_{\mathbf{\Theta}_{\mathcal{D}}} \overbrace{\frac{1}{N_\mathrm{slot} N^\mathrm{slot}_\mathrm{symb} N_A} \sum_{q=0}^{N_\mathrm{slot}-1} \sum_{\ell=0}^{N^\mathrm{slot}_\mathrm{symb}-1} \sum_{k=0}^{N_A-1} \left\lVert \left[ \mathbf{H} \right]_{q, \ell, k} - \left[ \mathbf{\check{H}} \right]_{q, \ell, k} \right\lVert_2^2}^{\mathcal{L}_\mathcal{D} \left(\mathbf{H},\mathbf{\check{H}}; \Theta_\mathcal{D} \right)}, 
\end{equation}

\noindent where $\mathcal{L}_\mathcal{D} \left(\mathbf{H},\mathbf{\check{H}}; \Theta_\mathcal{D} \right)$ is the loss function, and  $\left[ \mathbf{\check{H}} \right]_{q, \ell, k}$ is the compensated channel, which is given by eq. (\ref{eq:operation_compensator}). 
The neural compensator should work with a wide range of $\mathrm{SNR}^\mathrm{CSI}$ and aging $(\tau \in \mathbb{N}_0)$ value sto provide the neural precoder with a more accurate channel estimate.

\subsubsection{Solution Structure}

The design of the neural compensator takes the form of the DnCNN introduced in \cite{zhang2017beyond} for the problem domain of image denoising. The key idea behind the DnCNN approach is to estimate the noise (i.e., error) rather than attempting to estimate the corresponding noiseless image. Then, the estimated noise can be subtracted from the original input to produce an estimated representation of the noiseless image. As it is shown in \cite{zhang2017beyond}, this approach greatly outperforms the alternative of directly estimating of the noiseless image. 

We apply this approach to our particular problem by treating the channel matrix as the input image and its impairment as the noise to be removed. As it was mentioned in Section \ref{sec:5G advanced}, the term "impairment" refers to both aging and CE/CC error. According to our modeling abstraction, the Gaussian noise added to the available CSI information as per \eqref{eq:CSI_acq_model} models the UL CE error in TDD mode and  the DL CE error and CC error in FDD mode. It is important to note that the main difference between our work and existing works is that we treat channel aging in the coherence period as added noise, so that it can be estimated and removed. This is formally stated in the following proposition. 

\begin{proposition}[]
\label{prop:prediction_as_denoising}
Channel aging can be treated as noise; thus, channel prediction can be viewed as an instance of denoising. 
\begin{proof}
Using \eqref{eq:CSI_acq_model}, we can express a channel realization at slot $q$, i.e., $\left[ \mathbf{H} \right]_q$, from the previous channel realization at the CSI slot, $q_\mathrm{CSI} = f_\mathrm{CSI}(q)$, which was $\tau$ slots earlier, $f_\mathrm{CSI}(q) = q - \tau$, as
\begin{equation}
    \left[ \mathbf{H} \right]_q = \left[ \mathbf{H} \right]_{f_\mathrm{CSI}(q)} +  \left[ \mathbf{Z}^\mathrm{aging} \right]_{q},
\end{equation}

\noindent where $\left[ \mathbf{Z}^\mathrm{aging} \right]_{q}$ is the evolution of channel $\left[ \mathbf{H} \right]_{f_\mathrm{CSI}(q)}$ after a delay of $\tau$ slots:
\begin{equation}
    \label{eq:imparments}
    \left[ \mathbf{\tilde{H}} \right]_q = \left[ \mathbf{H} \right]_q - 
    \underbrace{\left[ \mathbf{Z} \right]^\mathrm{CSI}_{q}   - \left[ \mathbf{Z}^\mathrm{aging} \right]_{q}}_{\left[ \mathbf{Z}^\mathrm{imp} \right]_{q}},
\end{equation}

\noindent where $\left[ \mathbf{Z}^\mathrm{imp} \right]_{q}$ stands for the impairment (channel aging and CE/CC error) of the available CSI at slot $q$. 
Expression \eqref{eq:imparments} illustrates the fact that the impairments can be understood as an additive term to the actual channel at slot $q$, $\left[ \mathbf{\tilde{H}} \right]_q$. 

The available CSI at slot $q$ is the input of the neural compensator (i.e., denoiser) in order to remove the additive impairment and provides clean CSI as
\begin{align}
    \label{eq:compensator_as_denoiser}
    \left[ \mathbf{\check{H}} \right]_q &= \mathcal{D}^\mathrm{AI} \overbrace{ \left(\left[ \mathbf{H} \right]_{f_\mathrm{CSI}(q)} + \left[ \mathbf{Z}^\mathrm{CSI} \right]_{q} \right) }^{\left[ \mathbf{\tilde{H}} \right]_q}   \\
    &\overset{\mathrm{(a)}}{=} \left[ \mathbf{H} \right]_{f_\mathrm{CSI}(q)} + \left[ \mathbf{Z}^\mathrm{CSI} \right]_{q}  - \underbrace{\left(\left[ \mathbf{\hat{Z}}^\mathrm{CSI} \right]_{q}   - \left[ \mathbf{\hat{Z}}^\mathrm{aging} \right]_{q} \right)}_{\left[ \mathbf{\hat{Z}}^\mathrm{imp} \right]_{q}} \nonumber \\
    &\overset{\mathrm{(b)}}{=} \left[ \mathbf{H} \right]_q + \underbrace{\left(\left[ \mathbf{Z}^\mathrm{CSI} \right]_{q}   - \left[ \mathbf{Z}^\mathrm{aging} \right]_{q} \right)}_{\left[ \mathbf{Z}^\mathrm{imp} \right]_{q}}    - \left[ \mathbf{\hat{Z}}^\mathrm{imp} \right]_{q}, \nonumber
\end{align}

\noindent where (a) comes from the fact that the compensator's denoising architecture aims to estimate the additive impairment, $\left[ \mathbf{{Z}}^\mathrm{imp} \right]_{q}$ that contaminates the input tensor $\left[ \mathbf{\tilde{H}} \right]_q$ and (b) comes after isolating the term $\left[ \mathbf{H} \right]_{f_\mathrm{CSI}(q)}$ in \eqref{eq:CE_CC_H} and substituting it in (a). 

Finally, as it can be observed in \eqref{eq:compensator_as_denoiser}-(b), the output of the compensator, $\left[ \mathbf{\check{H}} \right]_q$, represents a \textit{prediction} of the true channel, $\left[ \mathbf{{H}} \right]_q$, which is based on an outdated and noisy channel sample that was taken $\tau$ slots earlier $\left(\left[ \mathbf{H} \right]_{f_\mathrm{CSI}(q)} + \left[ \mathbf{Z}^\mathrm{CSI} \right]_{q} \right)$. 
\end{proof}
\end{proposition}

As it can be observed in \eqref{eq:compensator_as_denoiser}-(b), the compensator tries to estimate the impairment $\left[ \mathbf{\hat{Z}}^\mathrm{imp} \right]_{q}$ in order to mitigate it. Ideally, a perfect estimate $\left[ \mathbf{\hat{Z}}^\mathrm{imp} \right]_{q}=\left[ \mathbf{{Z}}^\mathrm{imp} \right]_{q}$ would indicate the true channel is obtained for each slot $q$. 

\subsubsection{NN Architecture}
\begin{figure*}[t!]%
    \centering 
    \subfloat[]{{\includegraphics[width=0.12\textwidth, valign=c]{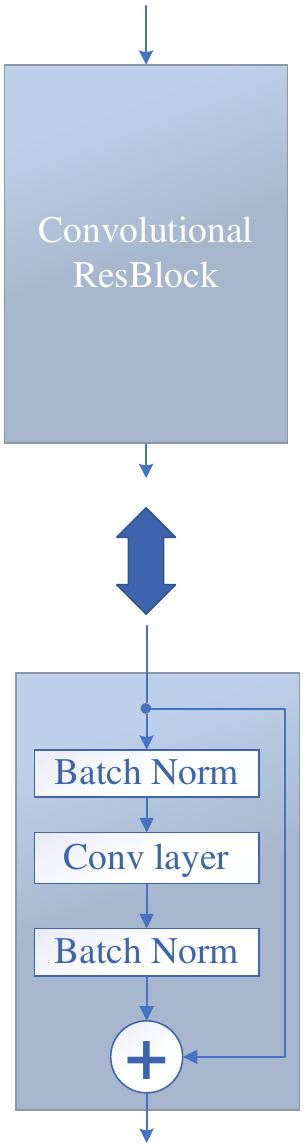} }}%
    \subfloat[]{{\includegraphics[width=0.28\textwidth, valign=c]{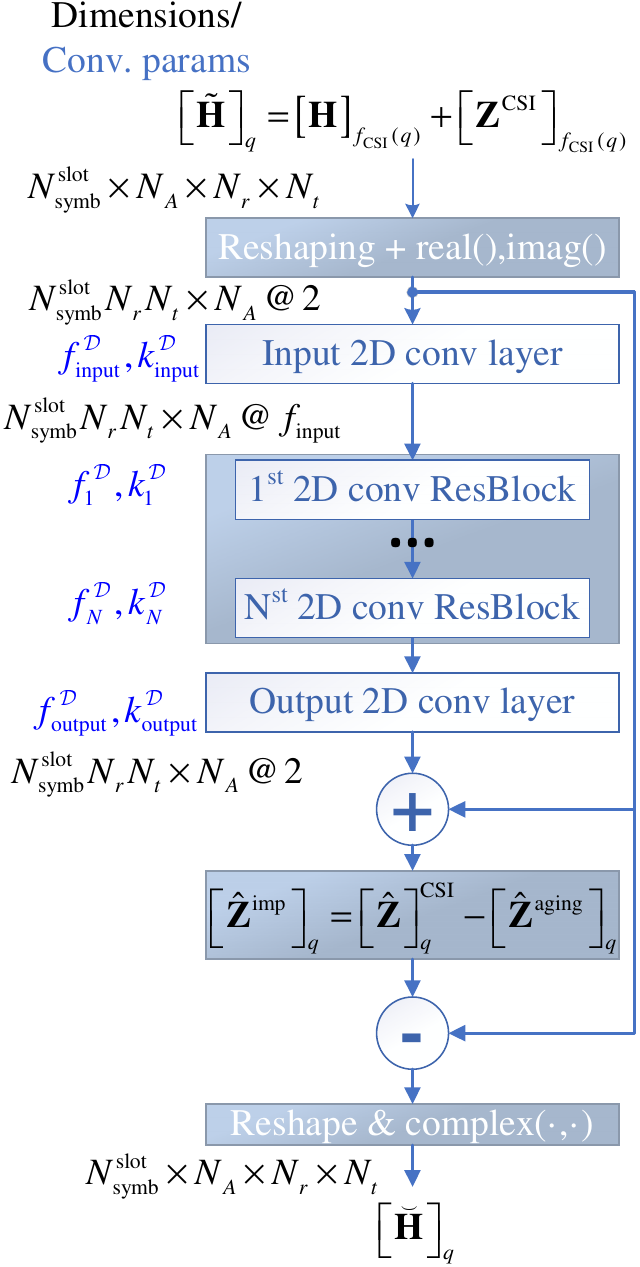} }}%
    \subfloat[]{{\includegraphics[width=0.28\textwidth, valign=c]{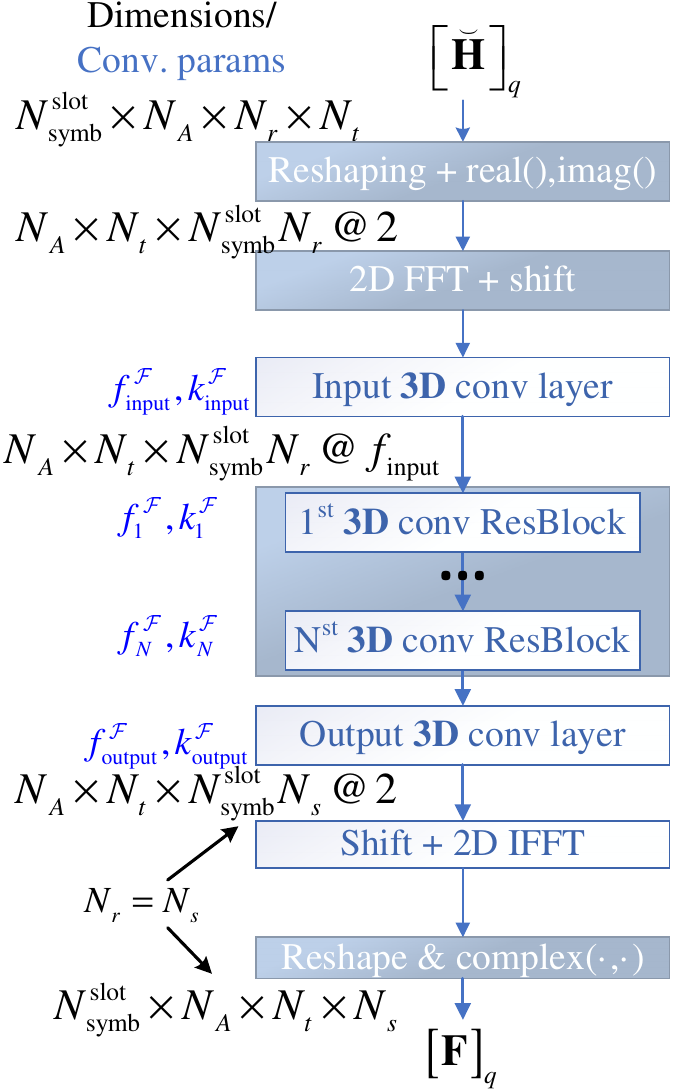} }}%
    \caption{NN architectures: (a) residual block; (b) proposed neural compensator to mitigate the CSI acquisition impairments, i.e., aging and CE/CC errors, based on \textbf{Proposition \ref{prop:prediction_as_denoising}}; and (c) proposed neural precoder based on 3D convolutional layers. The tensors' dimensions are written in black, whereas the CNN parameters (i.e., kernel size and number of filters) are written in blue.}%
    \label{fig:All_NNs}%
\end{figure*}

The neural compensator's architecture relies on \textbf{Proposition \ref{prop:prediction_as_denoising}} and thus applies the DnCNN's denoising concept to channel prediction, to mitigate both channel aging and CE/CC errors. The compensator's architecture is shown in Fig. \ref{fig:All_NNs}-(b) and comprised of the following parts:

\begin{itemize}
    \item \textit{Input layer}: The input data, $\left[ \mathbf{\tilde H} \right]_q \in \mathbb{C}^{N_\mathrm{symb}^\mathrm{slot} \times N_A \times N_r \times N_t}$, is reshaped to have the form $N_{{\rm{symb}}}^{{\rm{slot}}} N_r N_t  \times N_A @2$, where $@$ stands for the number of channels in the convolutional layers and represents the real and imaginary parts of $\left[ \mathbf{\tilde H} \right]_q$. Thus, after reshaping, the processed CSI has two dimensions and two channels. 
    \item \textit{Input 2D convolutional layer with no activation}: This layer has $f^\mathcal{C}_\mathrm{input}$ filters whose kernel size is $k^\mathcal{C}_{input}$.its size. The purpose of this first layer is to augment the size of the receptive field while expanding the input to be a feature tensor with the same number of channels as the subsequent RBs. After this layer, the output tensors all have the same shape, and the only change is in the number of channels, which depends on the number of filters in each layer. 
    \item \textit{Central part}: This part is comprised of $N$ RBs. As it is observed in Fig. \ref{fig:All_NNs}-(a), each residual block is composed of a \textit{batch normalization} operation, a 2D convolutional layer, a ReLU activation, and a residual connection the input of the RB. The $i$-th RB has $f^\mathcal{C}_{i}$ filters whose kernel size is $k^\mathcal{C}_{i}$.
    \item \textit{Output convolutional layer with no activation}: This layer has $f^\mathcal{C}_\mathrm{output} = 2$ filters whose size is $k^\mathcal{C}_\mathrm{output}$. The purpose of the output layer is to reconstruct the two input channels in order to estimate the impairment\footnote{In this architecture, the shape of the impairment tensor $\left[ \mathbf{\hat{Z}}^\mathrm{imp}\right]_q$, is $N_{{\rm{symb}}}^{{\rm{slot}}} N_r N_t  \times N_A @2$ instead of $N_\mathrm{symb}^\mathrm{slot} \times N_A \times N_r \times N_t$; however, we use the same symbol to represent both cases for the sake of notation simplicity.}, i.e., aging and CE/CC error, after a residual connection. 
    \item \textit{Denoising}: Once the impairment tensor $\left[ \mathbf{\hat{Z}}^\mathrm{imp}\right]_q$ has been estimated, it is removed from the input tensor, $\left[ \mathbf{\tilde H} \right]_q$ of shape $N_{{\rm{symb}}}^{{\rm{slot}}} N_r N_t  \times N_A @2$ to mitigate the impairment and clean up the available CSI. After denoising, the tensor is reshaped again to have the desired shape, $\left[ \mathbf{\check H} \right]_q \in \mathbb{C}^{N_\mathrm{symb}^\mathrm{slot} \times N_A \times N_r \times N_t}$.
\end{itemize}



It is worth noting that all the convolutional layers in this architecture are configured to add sufficient padding to avoid altering the tensors dimensions.

\subsection{Neural Precoder}
\subsubsection{Problem Formulation}
We aim to learn the set of parameters $\Theta_\mathcal{F}$ of the neural precoder 
$\mathcal{F}^\mathrm{AI} \left(\left[\mathbf{{H}} \right]_{q} \right)$ so that the precoding matrices $\left[ \mathbf{F} \right]_q$ minimize the BER with a range of SNR conditions. This problem is equivalent to maximizing the bit metric decoding (BMD) rate, which represents an achievable information rate for bit-interleaved coded modulation (BICM) systems \cite{2017Bocherer} and is expressed as

\begin{align}
\label{eq:BMD}
    \Gamma = 1 + \frac{1}{N_\mathrm{slot} N_s N_\mathrm{CB}} & \sum_{q = 0}^{N_\mathrm{slot}-1} \sum_{\nu = 0}^{N_s-1} \sum_{n = 0}^{N_\mathrm{CB}} 
        \Bigg( \left( 1 - \left[\mathbf{C} \right]_{q,\nu,n} \right)  \left[\mathbf{\hat{C}} \right]_{q,\nu,n} \nonumber \\
        &+ \log(1 + \mathrm{e}^{-\left[\mathbf{\hat{C}} \right]_{q,\nu,n}}) \Bigg),
\end{align}

\noindent where $\mathbf{C}$ represents the tensor of coded bits, and $\mathbf{\hat{C}} \in \mathbb{R}^{N_\mathrm{slot} \times N_s \times N_\mathrm{CB}}$ represents the tensor of LLR values, which can be understood as estimated logits for detected bits to be $1$ or $0$. Therefore the loss function of the neural precoder can be written as
$\mathcal{L}_\mathcal{F} \left( \mathbf{C}, \mathbf{\hat{C}}; \Theta_\mathcal{F}\right) = -\Gamma$. 

The fact that the neural precoder's input is the tensor of the CFR of all the subcarriers, $\left[\mathbf{H}\right]_q \in \mathbb{C}^{N_\mathrm{slot} \times N^\mathrm{slot}_\mathrm{symb} \times N_A \times N_r \times N_t}$, and the loss considers the LLRs values of all the subcarriers makes it possible to select wideband precoding matrices that minimize the BER of a whole transmission of $N_s$ layers, $N_A$ subcarriers and $N^\mathrm{slot}_\mathrm{symb}$ symbols. 
This enables us to directly apply the precoding matrices to all the subcarriers instead of having to rely on repeating single-carrier precoding methods for each subcarrier. It is worth noting that Sionna's fully differentiable approach makes it possible to use the BER as a target loss function to be minimized \cite{sionna}. 

\subsubsection{Solution structure}

Our solution is inspired by the super-resolution, fully convolutional approach proposed in \cite{soltani2019deep} in which a low-resolution version of a channel is mapped to its full-resolution counterpart. Unlike the now classic combination of 2D convolutional layers with fully connected layers, which employs a vectorization step, this fully convolutional approach avoids the vectorization step and instead leverages the existing data structure of the channel and precoding matrices to maintain the correlation present in the data, i.e., the correlation in the frequency dimension. 

The proposed 3D CNN architecture is modeled as a end-to-end mapping function $\mathcal{F}^\mathrm{AI} \left( \cdot \right)$ between the channel matrix tensor $\left[\mathbf{H} \right]_q \in \mathbb{C}^{N^\mathrm{slot}_\mathrm{symb} \times N_A \times N_r \times N_t}$ and its corresponding precoding tensor $\left[\mathbf{F}\right]_q \in \mathbb{C}^{N^\mathrm{slot}_\mathrm{symb} \times N_A \times N_t  \times N_s}$ as set out in \eqref{eq:operation_compensator}.

Without loss of generality, the frequency components of the channels $\left[\mathbf{H} \right]_{q,\ell,k}$ and  $\left[\mathbf{H} \right]_{q,\ell,k+1}$ are strongly correlated in the coherence bandwidth. This correlation is maintained after precoding, e.g., ZF or singular value decomposition (SVD), if pre-coding is completed on a per-carrier basis. The fully convolutional mapping function $\mathcal{F}^\mathrm{AI} \left( \cdot \right)$ is expected to maintain the correlation and make use of it in the outputted precoding matrices.

\subsubsection{NN Architecture}

Our proposed 3DCNN architecture is as shown in Fig \ref{fig:All_NNs}-(c): comprised of the following cascaded elements: 

\begin{enumerate}
    \item \textit{2D FFT module}: This element applies 2D FFT to the transmit antenna and frequency dimensions, i.e., $N_t$ and $N_A$, of the input channel with the same number of points as in the original dimension. This module is intended to transform the data to the angular-delay domain where information is more sparse. The same approach is used in \cite{wen2018deep} for massive MIMO CSI feedback. The FFT is followed by an FFT shift of $\floor*{\frac{N_\mathrm{FFT}}{2}}$ points in the frequency dimension. To center the main values, which would otherwise be placed at the edges of the dimension.

    \item \textit{Input layer}: This element is a 3D convolutional layer with linear activation and $f^\mathcal{F}_\mathrm{input}$ filters whose kernel is $k^\mathcal{F}_\mathrm{input} \times k^\mathcal{F}_\mathrm{input} \times k^\mathcal{F}_\mathrm{input}$.

    \item \textit{Stacked non-linear layers}: This element comprises $N$ stacked 3D convolutional layers with ReLU activation functions. The number of filters and the filter sizes are defined by the parameters $f^\mathcal{F}_{i}$ and $k^\mathcal{F}_{i}$, respectively, for $i \in [1, N] \subset \mathbb{N}$.

     \item \textit{Output layer}: This element is a 3D convolutional layer with $f^\mathcal{F}_\mathrm{output}$ filters of size $k^\mathcal{F}_\mathrm{output}$. It employs linear activation.

    \item \textit{2D IFFT module}: This element applies 2D IFFT to the transmit antenna and frequency dimensions, i.e., $N_t$ and $N_A$, of the output channel with the same number of points as in the original dimension. The IFFT is followed by a shift of $\floor*{\frac{N_\mathrm{FFT}}{2}}$ points in the frequency dimension. To undo the shift applied in in the 2D FFT module.

\end{enumerate}

\subsection{Training Methodology}
In this section, we describe in detail the procedure used to train the two sub-networks, which are fundamentally different. Both the training environments and the DL system simulation environment were implemented using Sionna \cite{sionna}, which is written in Python and relies on Keras and TensorFlow libraries. Among other features, Sionna supports on-the-fly sample generation rather than having to rely on pre-stored datasets. More importantly, Sionna treats communication blocks (such as those defined in Fig. \ref{fig:system}) as keras layers, which natively support backpropagation. This means that stochastic gradient descent (SGD) techniques can be used to minimize loss functions such as $\mathcal{L}_\mathcal{F}(\cdot)$.

\subsubsection{Neural Compensator}
The neural compensator was trained using conventional supervised learning. At each training iteration, Sionna generates a mini-batch of $N^\prime_\mathrm{slot}$ individual channels $\left[\mathbf{H}\right]_q$. Therefore, for a training process that consists of $N_{iterations}$, the overall number of slots is $N_\mathrm{slot} = N^\prime_\mathrm{slot} N_{iterations}$. At each iteration, the $\mathrm{SNR}^\mathrm{CSI}$, which models the CE/CC errors is randomly selected from a uniform distribution between $[\mathrm{SNR}^\mathrm{CSI}_\mathrm{min}, \mathrm{SNR}^\mathrm{CSI}_\mathrm{min}]$. Analogously, the channel aging, $\tau$ is randomly selected from a uniform distribution between $[\tau_\mathrm{min}, \tau_\mathrm{max}]$. 
Then, the ground truth is produced by applying $\tau$ slots of channel aging, and an error given by the $\mathrm{SNR}^\mathrm{CSI}$ to all the channels in the mini-batch of $N^\prime_\mathrm{slot}$ slots.

\subsubsection{Neural Precoder}
Sionna is used to minimize the loss function $\mathcal{L}_\mathcal{F} \left( \mathbf{C}, \mathbf{\hat{C}}; \Theta_\mathcal{F}\right) = -\Gamma$ from \eqref{eq:BMD}, via SGD. Its main characteristic is that it implements all modules using TensorFlow, which makes every node in the communication chain fully differentiable like as a NN. This makes Sionna able to plug an NN module into the system, compute a loss function with the desired metric at any point in the communication chain, and backpropagate the loss through all the elements in the communication chain all the way to the NN. All of this is done after each simulation step, which makes it completely unnecessary to extract and store a large fixed dataset for training. 

In each simulation run, $N_\mathrm{slot}$ independent slots are generated for propagation through the system and $N_\mathrm{slot}$ independent channel responses that last for $\tau N^\mathrm{slot}_\mathrm{symb}$ OFDM symbols are produced to account for channel aging. Also, in each simulation run, a random value of $\mathrm{SNR} \sim \mathcal{U}(\mathrm{SNR}_{\min}, \mathrm{SNR}_{\max})$ and applied as AWGN after the channel.

This novel implementation enables us to modify the last demapper stage to output LLR values for the data bits, convert the values to probabilities, and use the probabilities as part of the loss function. 

\subsubsection{Hyperparameter Tunning}

To find the architecture's optimal configuration, we performed a random search over a wide range of values for several parameters. More specifically, the parameters investigated were the number of non-linear layers $N$, the number of filters in each layer of the model $f_{[1, ..., N]| \mathrm{input}| \mathrm{output}}$, and the kernel size of each layer $k_{[1, ..., N]| \mathrm{input}| \mathrm{output}}$. Even though the kernels are 3D volumes, since we definethem as having equal-sized sides, their size is represented by only one number. 


The parameter configurations following the random search are summarized in Table \ref{tab:hyperparameter}.

\begin{table}[t]
\scriptsize
\centering
\caption{Selected hyperparameter for the neural compensator and the neural precoder}
\label{tab:hyperparameter}
\begin{tabular}{ c  c | c  c }
\toprule
\multicolumn{2}{c}{\textbf{Compensator}} & \multicolumn{2}{|c}{\textbf{Precoder}} \\
\hline 
\textbf{Parameter} & \textbf{Value} & \textbf{Parameter} & \textbf{Value}  \\  
\toprule
$N$ & $5$ & $N$ & $2$ \\
$\left(f^\mathcal{C}_\mathrm{input}, k^\mathcal{C}_\mathrm{input}\right)$ & $(96, 7)$ & $\left(f^\mathcal{F}_\mathrm{input}, k^\mathcal{F}_\mathrm{input}\right)$ & $(256, 9)$ \\
$\left(f^\mathcal{C}_{i}, k^\mathcal{C}_{i}\right)$ & $(96, 9)$ & $\left(f^\mathcal{F}_{i}, k^\mathcal{F}_{i}\right)$ & $(256, 7)$ \\
$\left(f^\mathcal{C}_\mathrm{output}, k^\mathcal{C}_\mathrm{output}\right)$ & $(96, 9)$ & $\left(f^\mathcal{C}_\mathrm{output}, k^\mathcal{C}_\mathrm{output}\right)$ & $(256, 5)$ \\
\bottomrule
\end{tabular}
\end{table}

\subsection{Complexity Analysis}
In this section, we analyze and compare the computational complexity, which is expressed in terms of the number of floating point operations (flops), of our baseline classical solution and our proposed AI-based solutions. 
The ZF algorithm considered in this work is based on the pseudo-inverse and computes $\left[\mathbf{\tilde{H}} \right]_{q,\ell, k}^{H}\left[\mathbf{\tilde{H}} \right]_{q,\ell, k}$ before computing its inverse in accordance with \eqref{eq:ZF_precoder}, which implies computing the eigenvalues and eigenvectors. Given that this operation needs to be performed for all $N_A$ subcarriers and all $N^\mathrm{slot}_\mathrm{symb}$ symbols, the resulting complexity is $\mathcal{O}(N_A \cdot N^\mathrm{slot}_\mathrm{symb} \cdot N_r^2 \cdot N_t)$.

The neural compensator, for its part, has an input convolutional layer, an output convolutional layer, and $N = 5$ RBs, each of which has a 2D convolutional layer. The time complexity of a standard convolutional layer without strides is $\mathcal{O}(f^2 \cdot k^2 \cdot H \cdot W)$,  where $f$ is the number of filters, $k$ is the size of the kernel, and $H$ and $W$ are the height and width of the input, respectively. If we take into account the fact that the padding was configured to keep the input's exact dimensions throughout the entire architecture and the NN parameters given with Table \ref{tab:hyperparameter}, we obtain the following computational complexity values:

\begin{itemize}
    \item Input: $\mathcal{O}(96^2 \cdot 7^2 \cdot N^\mathrm{slot}_\mathrm{symb} \cdot N_t \cdot N_r \cdot N_A)$
    \item Output: $\mathcal{O}(96^2 \cdot 9^2 \cdot N^\mathrm{slot}_\mathrm{symb} \cdot N_t \cdot N_r \cdot N_A)$
    \item RBs: $\mathcal{O}(5 \cdot 96^2 \cdot 9^2 \cdot N^\mathrm{slot}_\mathrm{symb} \cdot N_t \cdot N_r \cdot N_A)$
\end{itemize}

The neural precoder on the other hand, has 2D FFT and IFFT operations, input and output 3D convolutional layers, and $N = 2$ RBs. The time complexity of the 2D FFT and IFFT operations is $\mathcal{O}(H \cdot W \cdot \log(H) \cdot \log(W))$, where $H \times W$ are the dimensions of the input matrix. Assuming all three dimensions of the convolutional layers have the same kernel size and there are no strides, the time complexity of a 3D convolutional layer is given by $\mathcal{O}(f \cdot k^3 \cdot H \cdot W \cdot D)$, where $f$ is the number of filters, $k$ is the size of the kernel, and $H$, $W$ and $D$ are the height, width and depth of the input, respectively. If we take into account the fact that the padding was configured to keep the input's exact dimensions throughout the entire architecture and the NN parameters given with Table \ref{tab:hyperparameter}, we obtain the following computational complexity values:

\begin{itemize}
    \item Input: $\mathcal{O}(256 \cdot 9^3 \cdot N^\mathrm{slot}_\mathrm{symb} N_r \cdot N_t \cdot N_A)$
    \item Output: $\mathcal{O}(256 \cdot 5^3 \cdot N^\mathrm{slot}_\mathrm{symb} N_r \cdot N_t \cdot N_A)$
    \item Intermediate layers: $\mathcal{O}(2 \cdot 256 \cdot 7^3 \cdot N^\mathrm{slot}_\mathrm{symb} \cdot N_r \cdot N_r \cdot N_A)$
    \item 2D FFT/IFFT: $\mathcal{O}(2 \cdot N_A \cdot N_t \cdot \log(N_A) \cdot \log(N_t))$

\end{itemize}

It is worth noting that the convolution and FFT operations are highly optimized for GPU platforms, which can significantly reduce the computation time.

\section{Numerical Results}
\label{sec:Results}
In this section, we evaluate the performance of our proposed NN solution with a wide range of impairment levels and compare it to that of our baseline algorithms. 

\subsection{Training and Evaluation}

The system described in Section \ref{sec:System Model} was simulated with $N_t = 8$ transmit antennas, $N_r = 2$ receive antennas, and $N_s = 2$ data layers. The OFDM resource grid was designed to contain $N_\mathrm{FFT} = 32$ subcarriers with SCS of $\Delta f = 30$ kHz and $N^\mathrm{slot}_\mathrm{symb} = 14$ OFDM symbols per slot, where each slot lasts $0.5$ ms. Additionally, the simulation was set to produce QPSK modulation (i.e., $M=4$) with a code rate of $R = 0.5$. 
NN training and system performance evaluation was carried out using Sionna, Keras, and TensorFlow. 
%
%
Sionna's fully differentiable approach makes it possible to generate data samples on the fly rather than having to rely on a pre-generated dataset. The simulation was configured to generate $N^\prime_\mathrm{slot}$ independent channel realizations for each iteration in accordance with 3GPP's CDL-B model at a carrier frequency of $28$ GHz with a delay spread of $100$ ns and UE speeds randomly sampled between $3$ and $5$ km/h, which lead to Doppler frequencies, $f_{D}$, between $77.78$ Hz and $129.63$ Hz. 
The coherence time, $T_c$, ranges from $6.42$ ms to $3.85$ ms, i.e., from $13$ to $8$ slots, assuming a rough estimation of the coherence time as $T_c=0.5 c/(v f_c)$ \cite{2016Marzetta}, where $c$ is the speed of light, $v$ is the UE's speed in m/s and $f_c$ is the carrier frequency in Hz. 
%
%
Table \ref{tab:simparams} summarizes the main simulation parameters.

\begin{table}[t]
\scriptsize
\centering
\caption{Simulation parameters}
\label{tab:simparams}
\begin{tabular}{l l | l l}
\toprule
\textbf{Parameter} & \textbf{Value} & \textbf{Parameter} & \textbf{Value} \\ 
\toprule
$N_t$ & $8$ & Channel model & CDL-B     \\
$N_r$ & $2$ &  UE speed  (km/h) & $[3, 5]$     \\
$N_s$ & $2$ & Carrier frequency (GHz) & $28$     \\
$N^\mathrm{slot}_\mathrm{symb}$ &  $14$ & Delay Spread (ns) & $100$ \\ 
$N_\mathrm{FFT}$ & $32$  & $\Delta f$ (kHz) & $30$  \\
Pilot pattern & Kronecker & $\mathbf{L}$ & $(2, 11)$ \\
$\mathbf{k}_\mathrm{guard}$ & $(5, 6)$ & $\mathrm{SNR}^\mathrm{CSI}$ (dB) & $\{-5, 0, 5\}$ \\
$T_\mathrm{CSI}$ &  $11$ & $T_\mathrm{offset}$ & $0$ \\
$M$ & 4 & $R$ & $0.5$ \\ 
\bottomrule
\end{tabular}
\end{table}

\begin{table}[t]
\scriptsize
\centering
\caption{Training parameters for the neural compensator}
\label{tab:trainparams}
\begin{tabular}{ c  c | c  c }
\toprule
\multicolumn{2}{c}{\textbf{Compensator}} & \multicolumn{2}{c}{\textbf{Precoder}} \\
\hline
\textbf{Parameter} & \textbf{Value} & \textbf{Parameter} & \textbf{Value}  \\  
\toprule
$N_\mathrm{iterations}$ & $10^4$ & $N_\mathrm{iterations}$ & $10^4$ \\
$N^\prime_\mathrm{slot}$ & $32$ & $N^\prime_\mathrm{slot}$ & $64$ \\
$\mathrm{SNR}^\mathrm{CSI}_\mathrm{min}$ (dB) & $-5$ & $\mathrm{SNR}_\mathrm{min}$ (dB) & $-10$\\
$\mathrm{SNR}^\mathrm{CSI}_\mathrm{max}$ (dB) & $5$ & $\mathrm{SNR}_\mathrm{max}$ (dB) & $10$ \\
$\tau_\mathrm{min}$ & $0$ & - & - \\
$\tau_\mathrm{max}$ & $10$ & - & - \\
Learning Rate & 0.0001 & Learning Rate & 0.0001 \\ 
\hline
\end{tabular}
\label{tab:training_params_compensator}
\end{table}

\subsection{Channel Aging Compensation}

\begin{figure}[t!]
    \centering
    \includegraphics[width=0.8\columnwidth]{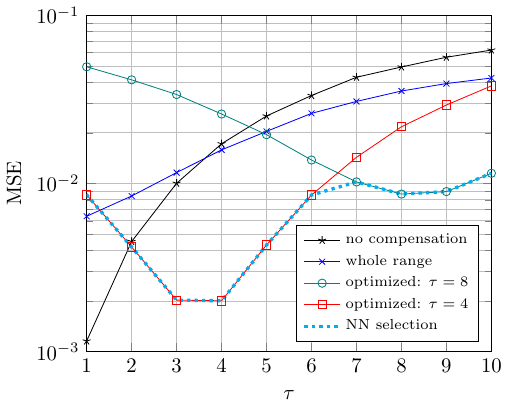}
    \caption{MSE performance of the neural compensator to mitigate channel aging.}
    \label{fig:MSE_aging}
 \end{figure}

In this section, we evaluate the neural compensator's ability to compensate for channel aging, which is expressed in terms of MSE between the available CSI at a given slot, $\left[\mathbf{\tilde{H}} \right]_q$ and the actual channel at that slot, $\left[\mathbf{{H}} \right]_q$. To assess solely the effect of aging, we do not consider CE/CC errors, i.e., $\mathrm{SNR}^\mathrm{CSI} \to \infty$, and thus, $\left[\mathbf{\tilde{H}} \right]_q = \left[\mathbf{H} \right]_{f_\mathrm{CSI}(q)}$. Fig. \ref{fig:MSE_aging} illustrates the MSE of different methods for different values of channel aging, $\tau$, which is expressed in terms of slots. 

When no compensation mechanism is used, channel aging severely degrades the available  CSI. This is reflected as an increase in the MSE as the black curve of Fig. \ref{fig:MSE_aging}.
To investigate the proposed solution's ability to compensate for channel aging, the proposed model was trained using different approaches, which are identified in Fig. \ref{fig:MSE_aging} as \textit{whole range}, \textit{optimized} (for $\tau = 4$ and $\tau = 8$ slots) and NN selection. These approaches are summarized as follows:
\begin{itemize}
    \item \textit{Whole range}: In this approach, the neural compensator is trained with aging values between $\tau_{min}$ and $\tau_{max}$ (see Table \ref{tab:training_params_compensator}). The goal here is to exploit the generalization abilities of the trained model for a wide range of aging values.  
    \item \textit{Optimized}: In this approach, the neural compensator is trained with only one aging value. This results in better performance for the optimization value chosen, at the cost of less generalization ability. Two NNs were optimized, one for $\tau = 4$ and one for $\tau = 8$ slots. 
    \item \textit{NN selection}: In this approach, we consider that the BS has two compensator models trained, one with $\tau = 4$ and one for $\tau = 8$. At each time slot $q$, the compensator that leads to smaller MSE is selected. 
\end{itemize}

In light of Fig. \ref{fig:MSE_aging} it can be concluded that the compensator can greatly reduce the impact of channel aging on the available CSI. This leads to a considerable improvement in end-to-end performance in terms of the BLER of data transmission, which we assess in the next section.




\subsection{Performance Results}

In this section, we evaluate the system performance in terms of BLER versus the DL SNR. We conducted two different experiments to accurately account for the compensation effect of both types of impairment considered, which are aging and Gaussian noise. In each experiment, a fixed value was assigned to one type of impairment in order to evaluate system performance at different levels of the remaining type of impairment. 


In the first experiment, the proposed solution and the baseline solution were evaluated with a fixed $\mathrm{SNR}^\mathrm{CSI}$ value of $5$ dB and a variety of aging values,  i.e., $\tau = (0,5, 10)$ slots. The slot duration in this case is $0.5$ ms, therefore with $v=70$ km/h, the coherence time of the channel is $8$ slots. 

\begin{figure}[t!]
    \centering
    \includegraphics[width=0.8\columnwidth]{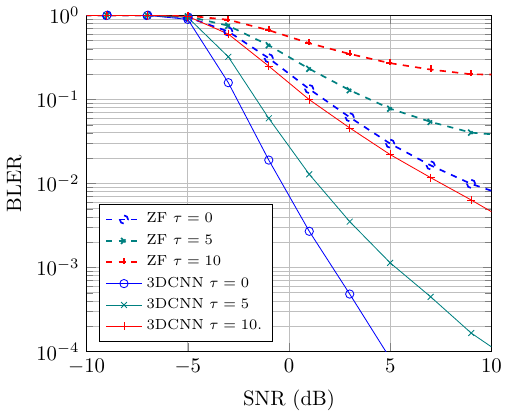}
    \caption{BLER results for different levels of aging and $\mathrm{SNR}^\mathrm{CSI} = 5$ dB.}
    \label{fig:bler_E10}
 \end{figure}

As it is shown in Fig. \ref{fig:bler_E10}, when $\tau = 0$ (i.e., no channel aging), our proposed solution exhibits around a $5.5$ dB increase in SNR for BLER values of $10^{-1}$ and a $9$ dB increase for BLER values of $10^{-2}$ compared to the baseline solution. In this case, the aging impairment is not present, and these gains can be attributed to Gaussian noise compensation and more accurate precoder computation of the proposed solution.

The gain in SNR that the proposed solution achieves over the baseline solution decreases for very small BLER values as the aging value, $\tau$, increases. Nevertheless, as the $\mathrm{SNR}$ increases, the ZF algorithm starts to exhibit an asymptotic behavior and is unable to achieve lower BLER values. When $\tau = 5$ slots and $\tau = 10 $ slots, the ZF algorithm presents asymptotes at $4 \cdot 10^{-2}$ and $2 \cdot 10^{-1}$, respectively. This is because at high DL $\mathrm{SNR}$ values, the impairment type that is applied dominates the non-idealities of the transmission. The impact that channel aging and CE/CC errors have on the available CSI impose an error floor on the of BLER of the baseline system that creates the need for the proposed AI-based impairment mitigation method.
In the same high $\mathrm{SNR}$ range, our proposed solution decreases BLER as the $\mathrm{SNR}$ increases with no visible asymptotic behavior. As expected, a high value of $\tau$ severely impacts the BLER performance of both algorithms; however, our proposed solution greatly outperforms the baseline. 

\begin{figure}[t!]
    \centering
    \includegraphics[width=0.8\columnwidth]{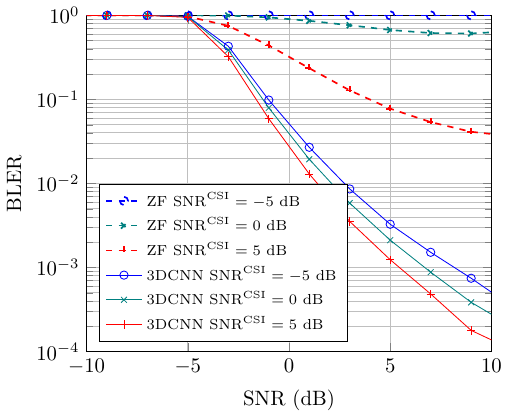}
    \caption{BLER results for different levels of CE/CC errors, $\mathrm{SNR}^\mathrm{CSI}$, and a fixed aging of $\tau = 5$ slots.}
    \label{fig:bler_E11_2.5ms}
\end{figure}

\begin{figure}[t!]
    \centering
    \includegraphics[width=0.8\columnwidth]{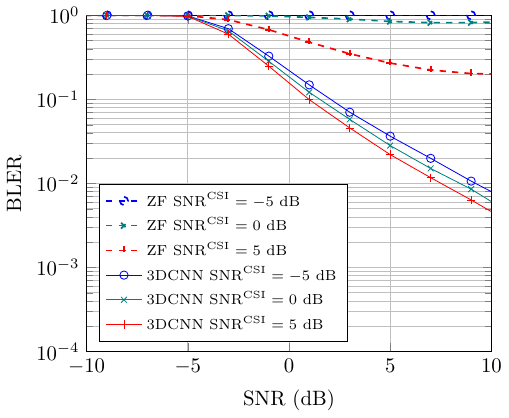}
    \caption{BLER results for different levels of CE/CC errors, $\mathrm{SNR}^\mathrm{CSI}$, and a fixed aging of $\tau = 10$ slots.}
    \label{fig:bler_E11_5ms}
\end{figure}

For the second experiment, the proposed solution and the baseline system were evaluated with various levels of Gaussian noise, $\mathrm{SNR}^\mathrm{CSI} = (-5,0,5)$ dB, and two values of aging, $\tau = (5,10)$. The results of these evaluations are illustrated in Fig. \ref{fig:bler_E11_2.5ms} and Fig. \ref{fig:bler_E11_5ms}, respectively.
The impairments are more severe in this second experiment, and thus, system performance is degraded even further. It can be observed that the baseline's performance is greatly reduced due to CE/CC errors. This is exacerbated with $\mathrm{SNR}^\mathrm{CSI} = 5$ dB, since the baseline system exhibits a BLER which is very close to $1$.
On the other hand, our proposed solution is very resilient to the same severe levels of impairment. As the value of $\mathrm{SNR}^\mathrm{CSI}$ decreases, the proposed solution's performance remains almost unaffected, exhibiting slight degradation between $1$ and $2.5$ dB.

\section{Conclusion}
\label{sec:Conclusions}
In this paper, we present a two-part NN solution for wideband precoding with CSI acquisition impairments, in the form of channel aging and CC/CE errors. To this end, we provide a generalized abstraction model that incorporates both the TDD and FDD CSI acquisition modes in 5G advanced. 
We show that our proposed solution's first sub-network, namely the neural compensator, is resilient to high levels of Gaussian noise and several OFDM frames of aging. This proves that, within the correlation time, channel aging can be predicted and subtracted the same way Gaussian noise can.
In our proposed solution's second sub-network, the neural precoder, we developed and trained a novel architecture to perform spatial multiplexing based on impaired CSI using 3DCNNs. The network's 3D structure makes it possible to successfully leverage the correlation that exists in the frequency domain of a MIMO channel by accounting for the unaltered frequency dimension in its design, which makes it suitable for and scalable to more extensive wideband signals. Unlike other NN solutions and classical algorithms, our proposed solution performs the precoding on all subcarriers without having to repeat the same operation for each subcarrier. 
Combining these two parts results in an accurate and resilient precoding NN that significantly outperforms the baseline solution considered, especially with severe levels of impairment. 
\vspace{-0.5cm}   

\bibliographystyle{IEEEtran}
\bibliography{refs.bib}

\end{document}